\pdfoutput = 1

\documentclass[11pt,preprint]{aastex}
\usepackage{latexsym}%
\usepackage{graphicx}%
\usepackage{amssymb}%
\usepackage{longtable}%
\usepackage{url}

\begin{document}

\title{A Targeted Search for Peculiarly Red L and T Dwarfs in SDSS, 2MASS, and WISE: Discovery of a Possible L7 Member of the TW Hydrae Association}
\author{Kendra Kellogg\altaffilmark{1,2,6}, Stanimir Metchev\altaffilmark{1,2}, Kerstin Gei$\ss$ler\altaffilmark{2}, Shannon Hicks\altaffilmark{2}, J. Davy Kirkpatrick\altaffilmark{3}, and Radostin Kurtev\altaffilmark{4,5}}
\altaffiltext{1}{Western University, Centre for Planetary and Space Exploration, 1151 Richmond St, London, ON N6A 3K7, Canada; kkellogg@uwo.ca, smetchev@uwo.ca}
\altaffiltext{2}{Stony Brook University, Stony Brook, NY 11790, USA}
\altaffiltext{3}{Infrared Processing and Analysis Center, Mail Code 100-22, California Institute of Technology, 1200 E. California Blvd., Pasadena, CA 91125, USA}
\altaffiltext{4}{Instituto de F\'{i}sica y Astronom\'{i}a, Facultad de Ciencias, Universidad de Valpara\'{i}so, Ave$.$ Gran Breta\~{n}a 1111, Playa Ancha, Casilla 53, Valpara\'{i}so, Chile}
\altaffiltext{5}{Millennium Institute of Astrophysics, Chile}

\begin{abstract}
We present first results from a targeted search for brown dwarfs with unusual red colors indicative of peculiar atmospheric characteristics. These include objects with low surface gravities or with unusual dust content or cloud properties. From a positional cross-match of SDSS, 2MASS and WISE, we have identified 40 candidate peculiar early L to early T dwarfs that are either new objects or have not been identified as peculiar through prior spectroscopy. Using low resolution spectra, we confirm that 10 of the candidates are either peculiar or potential L/T binaries. With a $J-K_s$ color of 2.62 $\pm$ 0.15 mag, one of the new objects --- the L7 dwarf 2MASS J11193254-1137466 --- is among the reddest field dwarfs currently known. Its proper motion and photometric parallax indicate that it is a possible member of the TW Hydrae moving group.  If confirmed, it would its lowest-mass (5--6 $M_{\rm Jup}$) free-floating member. We also report a new T dwarf, 2MASS J22153705+2110554, that was previously overlooked in the SDSS footprint. These new discoveries demonstrate that despite the considerable scrutiny already devoted to the SDSS and 2MASS surveys, our exploration of these data sets is not yet complete. 

\end{abstract}


\footnotetext[6]{Visiting Astronomer at the Infrared Telescope Facility, which is operated by the University of Hawaii under contract NNH14CK55B with the National Aeronautics and Space Administration.}

\maketitle

\section{Introduction}
Compared to main sequence stars, ultra-cool dwarfs display a wide range of near-infrared colors, even among objects at the same effective temperature or spectral type.  The diversity is diagnostic of the unique processes taking place in their molecule- and condensate-rich atmospheres.  Effective temperature is the main factor that governs the photospheric appearance of field-aged brown dwarfs, with current understanding pointing to a monotonic correspondence between effective temperature and optical spectral type \citep{vrba04, golim04, loop08}.  

\citet{cruz09} proposed a dimensional extension to the classification scheme for brown dwarfs, by incorporating surface gravity as a second parameter.  They adopt a qualitative description of surface gravities---intermediate, low, and very low---based on optical spectral line strengths.  \citet{allers13} expanded the classification scheme to the near-IR by adding continuum index measures to classify the absorption strengths of volatile molecules.

Low surface gravities generally contribute to higher dust content in the upper atmospheres of brown dwarfs, making them redder.  Analyses of the L and T dwarf population have shown that the optical and near-infrared colors of low-surface gravity objects are readily distinguishable from those of ``normal'' objects \citep[e.g.,][]{knapp04,cruz09,faher12,allers13}. However, there is also evidence of red brown dwarfs with high dust content without any signatures of youth \citep{loop08b, kirk10}. Their near-IR colors are very similar to those of the young, low-surface gravity objects but their spectra do not have any of the characteristics of youth. That is, peculiarly red brown dwarfs may not necessarily be low-gravity and hence young, but could instead be unusually dusty. As there have not been many unusually red old L dwarfs found, the cause of such dustiness is not well established. 

Finding the cause for the enhanced dust content is undoubtedly of interest for understanding the evolution of substellar objects, and the processes that affect the sedimentation and/or condensation of atmospheric dust.  It is also crucial for revealing the ages and properties of directly imaged extrasolar planets, most of which exhibit spectral energy distribution (SED) characteristics of both youth and high dust content \citep[e.g.,][]{mar12, bon13}. Because isolated brown dwarfs can be scrutinized much more readily than directly imaged extrasolar planets, we stand to potentially learn more about ultra-cool atmospheres from brown dwarfs than we can from exoplanets. 

Our understanding of the nature of brown dwarfs with unusual SEDs is presently hindered by the relatively small numbers of such peculiar objects.  Until recently, there have been no color-selected searches for peculiar brown dwarfs. Discoveries have been serendipitous, usually a by-product of searches for T dwarfs \citep[etc.]{loop08,loop08b,mcl07,burg04}. Only over the past few years have targeted searches been performed on large-area surveys to specifically seek unusually red objects (e.g., \citealp{aller13,gagne15}).

In view of this, we are conducting an independent program to purposefully seek L and T dwarfs with unusual optical/near-infrared (near-IR) colors.  The goal is to substantially expand the sample of peculiar L and T dwarfs in order to map the full range of their photospheric properties, and to better understand the evolution and content of their atmospheres.  We cross-correlated the SDSS, 2MASS, and WISE survey databases to seek candidate peculiar brown dwarfs based solely on photometric criteria.  Our first pass through the databases focused mainly on identifying unusually red objects.  Most notable among these is one of the reddest L dwarfs ever found (2MASS J11193254-1137466; 2MASS $J-K_s = 2.62 \pm 0.15$ mag). While peculiar L and T dwarfs have until now been found mostly serendipitously in large-scale photometric surveys, we have implemented a systematic approach to find these objects by design. We discuss the selection and prioritization of candidates in Section 2, and their follow-up observations in Section 3.  The spectroscopic characterization of the new L and T dwarfs is presented in Section 4.  In Section 5 we assess the significance of the findings from our systematic search of peculiar objects in the context of the presently known sample of L and T dwarfs.

\section{Candidate Selection}

We employ a photometric search for peculiar L and T dwarfs using combined optical (SDSS), near-IR (2MASS) and mid-IR (WISE) fluxes. Our candidate selection expands on the procedure presented in \cite{metchev08} and \cite{geissler11}, which applied joint positional and color constraints to search for T dwarfs in the overlap area of 2MASS and SDSS DR1 (2099 deg$^{2}$). We use the ninth Data Release (DR9) from SDSS \citep{ahn12}, which has a 14555 deg$^{2}$ footprint, encompassing an area 6.9 times larger than the DR1 footprint. The $>$10-year observational epoch difference between 2MASS and SDSS DR9 prompts us to choose a much larger cross-match radius than was used in the first two studies. We use the Virtual Astronomical Observatory catalog cross-comparison tool\footnote{\url{http://vao-web.ipac.caltech.edu/applications/VAOSCC/}} and chose a cross-match radius of $16\farcs5$ to maintain sensitivity to objects with proper motions as high as $1\farcs5$ yr$^{-1}$.

\subsection{Selection Criteria
\label{sec:selection}}

Our magnitude and color selection criteria are summarized below. In the following, all $\it{griz}$ magnitudes are on the SDSS photometric scale \citep{lup02}, and the 2MASS and WISE magnitudes are on the Vega scale:
\begin{enumerate} \itemsep0pt \parskip0pt
\item $z-J$ $>$ 2.5 mag;
\item $i-z$ $>$ 1.5 mag;
\item $J$ $>$ 14 mag;
\item $z$ $\leq$ 21 mag and $z_{err}$ $\leq$ 0.2 mag;
\item no $g, r <$ 23 mag detection within $1\farcs3$ of the 2MASS coordinate;
\item SDSS object flag setting type = 6 or 3 (point or extended source);
\item 2MASS object flag setting mp\_flg = 0 (i.e., not marked as a known minor planet), gal\_contam = 0 (i.e., not contaminated by a nearby 2MASS extended source), and ext\_key = NULL (i.e., not extended in 2MASS);
\item $H-W2$ $>$ 1.2 mag;
\item $z-J > -0.75(J-K_{s})+ 3.8$ mag (criterion used only to prioritize follow-up of red outliers).
\end{enumerate}

Our $z-J$ and $i-z$ color cuts (criteria 1 and 2) were chosen to ensure sensitivity to L and T dwarfs, all of which have a steep red optical slope. The $J$ $>$ 14 magnitude cutoff was imposed to minimize the large number of candidates representing the cross-identification of a bright star artifact in SDSS (e.g., a filter glint or a diffraction spike, especially near saturated stars) with the (unsaturated) image of the same star in 2MASS. Criterion 4 was chosen to ensure detection in SDSS with at least a moderate SNR. 

Our $16\farcs5$ matching radius commonly resulted in multiple matches of nearby faint SDSS objects to the same, brighter 2MASS object.  Each of these individual matches would nominally satisfy the color and magnitude selection criteria, since the faint SDSS photometry would be paired with the brighter 2MASS photometry.  However, visual inspection clearly demonstrated that the SDSS and 2MASS objects were distinct, and that the actual object in SDSS that positionally matched the 2MASS object was not nearly as red, and so did not satisfy the $z-J>2.5$~mag criterion.  Therefore, we discarded any object that had a $g$-band detection (i.e. $g \leq$ 23 mag and likely a star) in the SDSS catalog within $1\farcs3$ (the angular resolution of SDSS) of the original 2MASS coordinates (criterion 5). This removed $\sim$86\% of the candidate sample.

The SDSS object flag restrictions (criterion 6) ensure that the identified candidates are not known artifacts or flux measurements of the blank sky in SDSS. The SDSS morphological star-galaxy separation is $<$ 97\% accurate for $r$ $\geq$ 21 mag \citep{yas01} so we include both star and galaxy object types in this criterion in case our faint brown dwarfs were mis-classified. We also wanted to ensure that they are not known minor planets, extended or contaminated by nearby extended sources in 2MASS (criterion 7).

To make sure that all of the objects in our candidate list were real objects, we cross-matched our list with the WISE All-Sky Data Release using the SDSS coordinates. Our objects are expected to be detected in the WISE $W1$ band because outside of the galactic plane the $W1$ SNR $=5$ level corresponds to $\lesssim$ 16.8~mag.\footnote{\url{http://wise2.ipac.caltech.edu/docs/release/allsky/expsup/sec6_3a.html}} This matches the 2MASS $K_{s}$ flux limit at high galactic latitude, especially since L and T dwarfs have positive $K_{s}-W1$ colors.  A radius of $16\farcs5$ was again chosen for this cross-match.  An additional color cut was applied on $H-W2$ (criterion 8) in order to select only L and T dwarfs (based the color-spectral type relations from \citealp{kirk11}). This removed $\sim$74\% of the remaining sample.

Finally, we visually inspected the images of remaining candidates using the Infrared Science Archive Finder Chart service\footnote{\url{http://irsa.ipac.caltech.edu/applications/FinderChart/}} and removed objects that were contaminated by nearby extended sources in SDSS. This eliminated approximately 22\% of the remaining candidate sample leaving us with 314 candidates (Figure \ref{fig:ccdiag_paper}a).

\subsection{Prioritization of Peculiar Objects}

\begin{figure} 
\centering 
\includegraphics[scale=0.45]{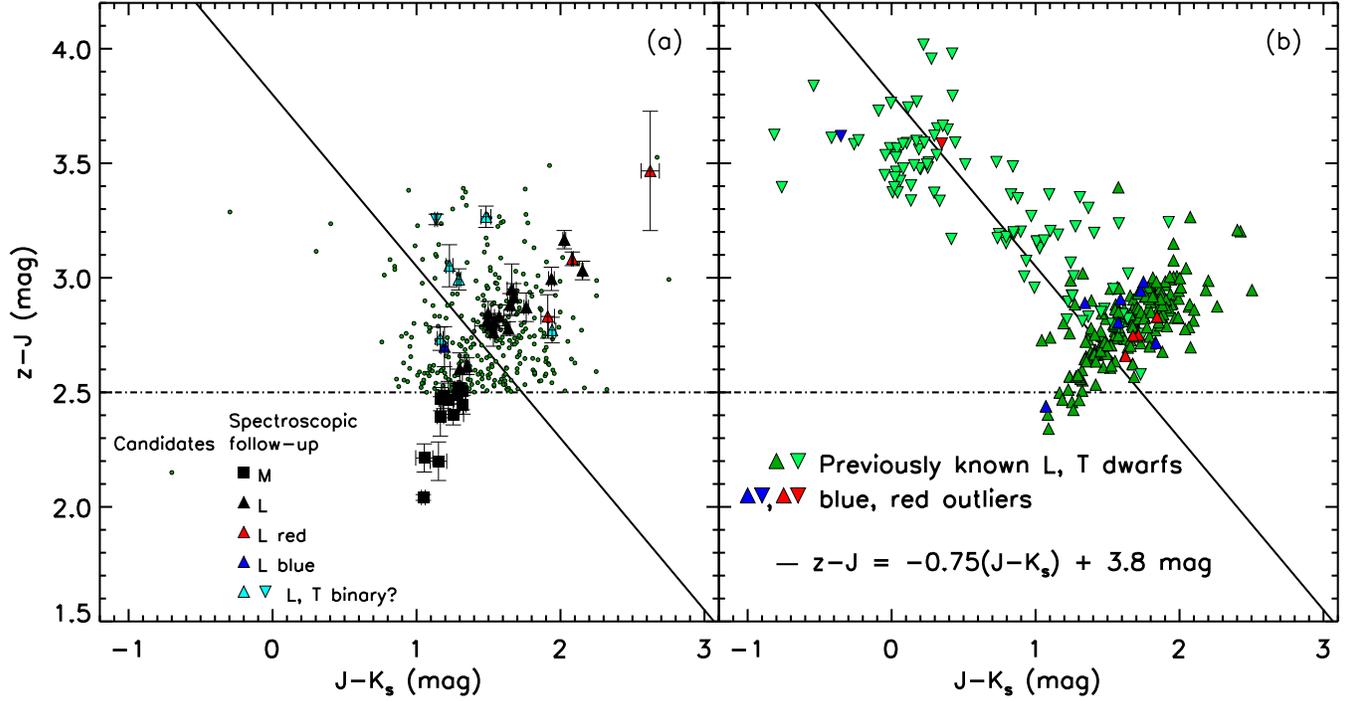}
\caption{ (a) Photometric color-color diagram of all L and T dwarf candidates redder than $z-J$ = 2.5 mag (green dots) identified in our SDSS-2MASS-WISE cross-match. All other symbols (squares - M dwarfs; upwards triangles - L dwarfs; downwards triangle - T dwarf) represent the synthetic colors of the candidates followed up with spectroscopic observations so far. The black symbols are ``normal" objects and the red and blue symbols are objects that we have identified as peculiar or binary. Objects redder than the $z-J=-0.75(J-K_s)+3.8$~mag line are candidate peculiarly red L and T dwarfs and were prioritized for spectroscopic follow-up. (b) SDSS/2MASS synthetic color-color diagram of L and T dwarfs from the SpeX Prism Archive (upwards and downwards triangles, respectively). The $z-J$, and $J-K_s$ colors were formed synthetically from the SpeX spectra. Two-sigma red and blue photometric color outliers are indicated by red and blue symbols, respectively. The $z-J=-0.75(J-K_s)+3.8$~mag line was designed to select the photometric red outliers.}
\label{fig:ccdiag_paper}
\end{figure}

Since our goal was to select unusually red brown dwarfs in the absence of spectral type information, an additional color criterion (9) was set in order to prioritize ed objects.  To decide the form of the color criterion, we first analyzed the spectra of L and T dwarfs in the SpeX Prism Archive\footnote{\url{http://pono.ucsd.edu/~adam/browndwarfs/spexprism/}} by forming synthetic photometry over various red-optical and near-IR bandpasses. These L and T dwarfs with archival SpeX data formed our control sample, based on which we designed our $z-J$ vs. $J-K_s$ criterion 9 (Figure \ref{fig:ccdiag_paper}b).

Given available spectral type information, the unusually red objects in the control sample were set to be those for which the $J-K_s$ color was $>$2-$\sigma$ redder than the median for the spectral subtype.  The medians and standard deviations of the $J-K_s$ colors of M8-T8 dwarfs were adopted from Faherty et al. (2009; M8-M9 and T0-T9) and from Faherty et al. (2013; L0-L9).  The unusually red objects in the control sample are shown with red symbols in Figure \ref{fig:ccdiag_paper}b. The number of objects from our sample that passed this criteria was 178.

The color prioritization did not streamline our observational follow-up strategy significantly, as the scatter in colors among spectral types is larger than the scatter at any given spectral type.  Nonetheless, we did observe the reddest candidates whenever possible, and included observations of lower-priority targets only as necessary.

\section{Spectroscopic Observations and Data Reduction}

Once our candidates were selected, we performed follow-up spectroscopic observations of 40 of the objects ($\sim$13\% of the total candidate sample; 22 high priority and 18 lower priority) using the SpeX instrument \citep{rayner03} on the NASA Infrared Telescope Facility (IRTF) and the Folded-port InfraRed Echellette (FIRE) instrument \citep{simcoe08} on the Magellan Baade telescope. Conditions were photometric on most nights, except on August 3, 2011, April 18, 2012, April 19, 2012, and July 14, 2012, when there was scattered cirrus. All reduction of the low-resolution spectra (SpeX and FIRE LD) was done in Interactive Data Language (IDL). 

\subsection{IRTF/SpeX}
The majority of our follow-up observations were taken using the SpeX spectrograph on the IRTF. The broad, simultaneous wavelength coverage (0.8--2.5 $\mu$m) of SpeX and its location in the northern hemisphere are ideal for follow-up of SDSS-identified candidates. These spectra were obtained between 2011 August and 2013 June. The observations were taken in prism mode either with the $0\farcs8 \times 15\farcs0$ or with the $1\farcs6 \times 15\farcs0$ slit, resulting in resolutions of $R \sim$150 and $\sim$75, respectively. The slit orientation was maintained to within 20$^{\circ}$ of the parallactic angle for all targets. We used a standard A-B-B-A nodding sequence along the slit to record object and sky spectra. Individual exposure times were either 60 s or 180 s per pointing. The shorter exposure times allowed us to better subtract the sky-glow under changing atmospheric conditions. Standard stars were used for flux calibration and telluric correction. Flat-field and argon lamps were observed immediately after each set of target and standard star observations for use in instrumental calibrations. Observation epochs and instrument settings for each science target are given in Table \ref{tab:spex}.

All reductions of the data taken with SpeX were carried out with the {\sc spextool} package version 3.4 \citep{cushing04, vacca03}, using a weighted profile extraction approach \citep{horne86, rob86}. The aperture widths were set to be the radius at which the spatial profile dropped to $\sim$5\% of the peak flux value to ensure no contamination from background noise; the background regions were chosen to begin at the edge of the PSF radius (i.e., beyond 2.5 pixels = $0\farcs 375$). A constant value was fit to the background and subtracted from the spectrum. The individual extracted and wavelength calibrated spectra from a given sequence of observations, each with their own A0 standard, were then scaled to a common median flux and median-combined using {\sc x\_combspec}.  The combined spectra were corrected for telluric absorption and flux-calibrated using the respective telluric standards with {\sc x\_tellcor}. All calibrated sets of observations of a given object were median combined to produce the final spectrum. The reduced spectra were smoothed to the instrumental resolution corresponding to the chosen slit width, using the Savitzky-Golay smoothing kernel \citep{press92}.

\begin{deluxetable}{ccccccc}
\tabletypesize{\small}
\tablecolumns{6}
\tablewidth{0pt}
\tablecaption{SpeX Observations
\label{tab:spex}}
\tablehead{
\colhead{Identifier} & \colhead{Date} & \colhead{2MASS J} & \colhead{Slit Width} & \colhead{Exposure} & \colhead{A0 Calibrator} \\
\colhead{(J2000)} & \colhead{(UT)} & \colhead{(mag)} & \colhead{(arcsec)} & \colhead{(min)} & \colhead{$ $} }
\startdata
\object{2MASS J08095903+4434216} & 2011 Dec 31 & 16.44 & 0.8 & 24 & HD 75135\\
\object{2MASS J09572983+4624177} & 2013 Jun 06 & 16.25 & 1.6 & 24 & HIP 53735\\
\object{2MASS J10020752+1358556} & 2013 Jun 07 & 17.19 & 1.6 & 32 & HIP 35735, HIP 54815\\
\object{2MASS J11193254-1137466} & 2013 Jun 06 & 17.29 & 1.6 & 16 & HIP 53735\\
... \tablenotemark{a} & 2013 Jun 07 & 17.29 & 1.6 & 8 & HIP 54815\\
\object{2MASS J11260310+4819256} & 2013 Jun 07 & 17.20 & 1.6 & 24 & HIP 54815, HIP 56147\\
\object{2MASS J13043568+1542521} & 2013 Jun 06 & 17.32 & 1.6 & 70 & HIP 68209, HIP 68868\\
\object{2MASS J13431670+3945087} & 2011 Dec 31 & 16.16 & 0.8 & 16 & HD 125798\\
\object{2MASS J14025564+0800553} & 2013 Jun 07 & 16.84 & 1.6 & 160 & HIP 68868, HIP 116886\\
\object{2MASS J16005759+3021571} & 2011 Aug 02 & 16.97 & 0.8 & 54 & HD 153650\\
\object{2MASS J16094569+1426422} & 2011 Aug 03 & 16.84 & 0.8 & 60 & HD 152531\\
\object{2MASS J16091143+2116584} & 2011 Aug 02 & 16.96 & 0.8 & 60 & HD 153650\\
\object{2MASS J16135698+4019158} & 2012 Apr 19 & 17.05 & 0.8 & 48 & HD 151353\\
\object{2MASS J16231308+3950419} & 2012 Apr 18 & 16.97 & 0.8 & 60 & HD 165623\\
\object{2MASS J16242936+1251451} & 2011 Aug 03 & 16.26 & 0.8 & 36 & HD 152531\\
\object{2MASS J16304999+0051010} & 2012 Jul 14 & 16.00 & 0.8 & 12 & HD 157359\\
\object{2MASS J16322360+2839567} & 2012 Apr 18 & 16.63 & 0.8 & 48 & HD 158261\\
\object{2MASS J16360752+2336011} & 2012 Jul 14 & 16.88 & 0.8 & 12 & HD 157359\\
\object{2MASS J16370238+2520386} & 2011 Aug 03 & 16.50 & 0.8 & 36 & HD 157359\\
\object{2MASS J16403870+5215505} & 2012 Apr 19 & 17.22 & 0.8 & 60 & HD 155838\\
\object{2MASS J16410015+1335591} & 2011 Aug 03 & 16.90 & 0.8 & 48 & HD 157359\\
\object{2MASS J16470847+5120088} & 2012 Apr 19 & 17.03 & 0.8 & 48 & HD 155838\\
\object{2MASS J16592987+2055298} & 2012 Jul 15 & 16.33 & 1.6 & 90 & HD 164728\\
\object{2MASS J17081563+2557474} & 2011 Aug 02 & 16.42 & 0.8 & 48 & HD 164728\\
\object{2MASS J17145224+2439024} & 2012 Jul 14 & 16.84 & 0.8 & 12 & HD 165623\\
\object{2MASS J17161258+4125143} & 2011 Aug 03 & 16.75 & 0.8 & 36 & HD 165623, HD 165622\\
 ... \tablenotemark{a} & 2012 Jul 15 & 16.75 & 1.6 & 36 & HD 165623\\
\object{2MASS J17251557+6405005} & 2012 Jul 15 & 16.81 & 1.6 & 24 & HD 165622\\
\object{2MASS J17373467+5953434} & 2012 Apr 19 & 16.88 & 0.8 & 60 & HD 166639\\
\object{2MASS J21050130-0533505} & 2011 Aug 03 & 16.42 & 0.8 & 36 & HD 209051\\
\object{2MASS J21111559-0543437} & 2011 Aug 03 & 16.09 & 0.8 & 36 & HD 209051\\
\object{2MASS J21115335-0644172} & 2011 Aug 02 & 16.90 & 0.8 & 60 & HD 209051\\
\object{2MASS J21203483-0747378} & 2011 Aug 02 & 16.82 & 0.8 & 60 & HD 210253\\
\object{2MASS J21243864+1849263} & 2013 Jun 06 & 17.03 & 1.6 & 56 & HIP 53735, HIP 68209, HIP 68868\\
\object{2MASS J21392224+1124323} & 2011 Aug 03 & 16.49 & 0.8 & 48 & HD 210265\\
 ... \tablenotemark{a} & 2011 Dec 31 & 16.49 & 0.8 & 36 & HD 210265\\
 ... \tablenotemark{a} & 2012 Jul 15 & 16.49 & 0.8 & 48 & HD 210265\\
\object{2MASS J22153705+2110554} & 2013 Jun 07 & 16.00 & 1.6 & 72 & HIP 116886\\
\object{2MASS J22483513+1301453} & 2011 Dec 31 & 16.82 & 0.8 & 28 & HD 220184\\
 ... \tablenotemark{a} & 2012 Jul 14 & 16.82 & 1.6 & 36 & HD 220184\\
 ... \tablenotemark{a} & 2012 Jul 15 & 16.82 & 0.8 & 24 & HD 210265\\
\object{2MASS J23023319-0935188} & 2012 Jul 15 & 16.80 & 0.8 & 60 & HD 222903\\
\object{2MASS J23322678+1234530} & 2013 Jun 06 & 16.89 & 1.6 & 42 & HIP 68868\\
\object{2MASS J23443744-0855075} & 2011 Aug 02 & 16.77 & 0.8 & 60 & HD 2717\\
\enddata
\tablenotetext{a}{\footnotesize Repeat observations of the same object, combined with the previous data.}
\end{deluxetable}

\subsection{Magellan/FIRE}
Two of the 40 total candidates were observed using the FIRE spectrograph on the 6.5 m Magellan telescope. The observations of these objects were taken in the low-dispersion (LD) mode with the $0\farcs6 \times 50\farcs0$ longslit resulting in a resolution of $\sim$400. We used a standard A-B-B-A nodding sequence along the slit to record object and sky spectra. Individual exposure times ranged from 31.7--126.8 s per pointing, depending on the brightness of the object. Standard stars were used for flux calibration and telluric correction. We used optimal gain settings of 1.2 e$^{-}$/DN and 3.8 e$^{-}$/DN for the science targets and 3.8 e$^{-}$/DN for the standards as suggested in the FIRE observing manual\footnote{\url{http://web.mit.edu/~rsimcoe/www/FIRE/ob_manual.htm}}. Illumination and appropriate pixel flats were observed either at the beginning or the end of the night and a neon-argon lamp was observed immediately after each set of target and standard star observations for use in instrumental calibrations. All science and telluric observations were taken using the sample-up-the-ramp (SUTR) readout mode whereas all calibration observations were taken in Fowler 1 mode due to the shortness of exposure times. Observation epochs and instrument settings for each target are given in Table \ref{tab:fire}.

\begin{deluxetable}{ccccccccc}
\tablecolumns{8}
\tablewidth{0pt}
\tabletypesize{\small}
\tablecaption{FIRE Observations
\label{tab:fire}}
\tablehead{
\colhead{2MASS ID} & \colhead{Date} & \colhead{J} & \colhead{Dispersion} & \colhead{Slit Width} & \colhead{Gain} & \colhead{Exposure} & \colhead{A0 Calibrator} \\
\colhead{(J2000)} & \colhead{(UT)} & \colhead{(mag)} & \colhead{Mode} & \colhead{(arcsec)} & \colhead{(e$^{-}$/DN)} & \colhead{(min)} & \colhead{}}
\startdata
\object{2MASS J07483864+1743329} & 2012 Mar 21 & 16.27 & Long Slit & 0.60 & 3.8 & 4.2 & HD 57450\\
\object{2MASS J16110632+0025469} & 2012 Mar 21 & 17.02 & Long Slit & 0.60 & 1.2 & 16.9 & HD 153940\\
\enddata
\end{deluxetable}

The FIRE low-dispersion spectra were reduced using the FIREHOSE Low Dispersion package which evolved from the optical echelle reduction software package MASE \citep{boch09}. The spectra were extracted using the optimal extraction approach with the aperture radius being the PSF radius (usually $\sim$3 pixels = 0$\farcs$45) which was then masked to prevent biasing to the sky model. A local background was modeled using a basis spline (i.e., piecewise polynomial) fit to the masked profile and subtracted from the spectra which were subsequently extracted using a weighted profile extraction approach \citep{horne86}. The extracted spectra were wavelength-calibrated and each set of observations were median-combined. The combined spectra were corrected for telluric absorption and flux-calibrated with their associated A0 calibration star. All calibrated sets of observing sequences of a given object were median combined to produce a final spectrum. The reduced spectra were smoothed, using the IDL Savitzky-Golay smoothing algorithm, to the same resolution as the SpeX standards  for comparison.

\subsection{Synthetic Photometry
\label{sec:synphot}}

While comparing the 2MASS colors of our L and T dwarf candidates to their spectra, we noticed that in a significant fraction of cases the 2MASS colors were too red compared to the spectra.  All of our objects were flux-calibrated with A0 stars with known $B-V$ colors, observed at similar airmasses, so we had no reason to suspect a chromatic effect in our flux calibration.  Instead, the reason for the discrepancy was traced to flux over-estimation bias at low SNR in 2MASS. 

\begin{figure}[h]
\centering
\includegraphics[scale=0.5]{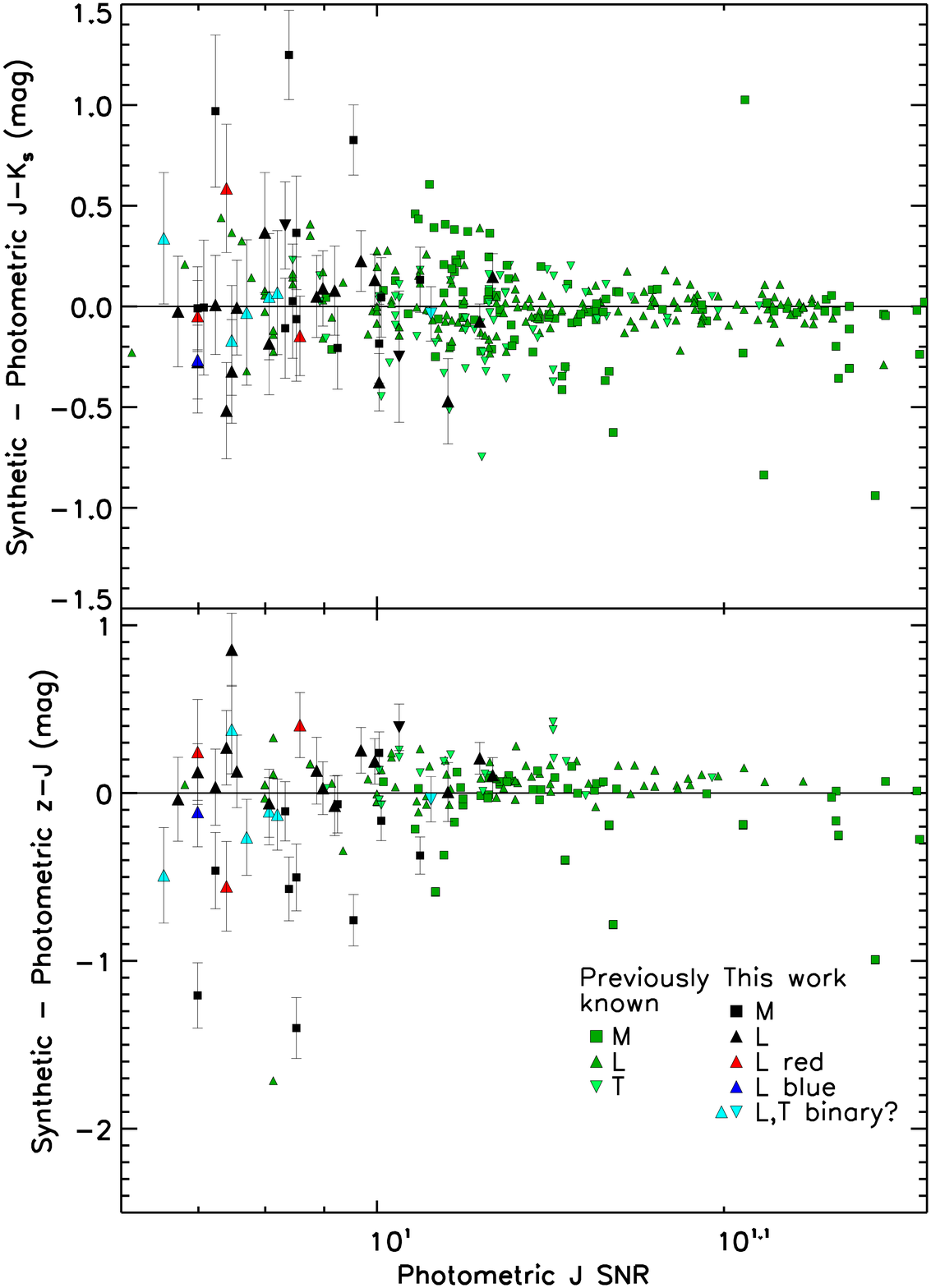}
\caption{Difference in synthetic vs photometric $J-K_s$ and $z-J$ colors for M, L and T dwarfs from the SpeX Prism Archive (green symbols) and for objects from this work (all other colored symbols). The black symbols are ``normal" objects and the blue and red symbols are objects that we have identified as peculiar or binary. Fewer objects appear in the $z-J$ comparison figure (lower panel) because not all SpeX Archive objects are in the SDSS database.}
\label{fig:colorcompare}
\end{figure}

Our objects are faint and often near the SNR = 5 detection limit of 2MASS in the {\it J}-band filter.  The greater noise near the detection limit means that objects that would normally be below the limit have a finite chance of appearing brighter because of statistical variations.  The effect enhances the number of faint objects with low SNR in a flux-limited survey, becoming increasingly important at SNR $<10$ \citep{bias}.  Because all of our objects are faint and red, their 2MASS $J$-band magnitudes preferentially suffer from this bias, resulting in redder than expected $z-J$ colors. This effect is particularly large in the case of the few faint M-dwarfs that entered our sample because of their biased photometric colors (section 4.1). Figure \ref{fig:colorcompare} shows how the synthetic colors compare to the photometric colors as a function of the photometric $J$-band SNR for both $z-J$ and $J-K_s$. Indeed, at lower SNR, the $z-J$ photometric colors are on average redder than their synthetic colors while the $J-K_s$ photometric colors are on average slightly bluer.

For the remainder of our analysis we use only synthetic SDSS $z$ and 2MASS $JHK_s$ magnitudes for our candidates and for previously known objects with SpeX Prism Archive spectra. The errors on the synthetic photometry in Table \ref{tab:results} are standard errors derived from the scatter among the continuum slopes of the individual 60 s or 180 s-exposure spectra of our targets and their corresponding standard stars.  These errors incorporate systematic uncertainties from potential chromatic slit losses should the targets have been imperfectly positioned on the slit.

\section{Spectral Classification Results}
We estimate spectral types for our objects by comparing them to spectra of brown dwarfs available from the SpeX Prism Archive\footnote{\cite{kirk10}, \cite{burg10}, \cite{burg08}, \cite{burg07b}, \cite{loop07}, \cite{burg06}, \cite{chiu06}, \cite{reid06}, and \cite{burg04}}.  When our spectra don't match any of the normal brown dwarf spectra, we compare to other unusual spectra. In this way, we are able to assess potential spectroscopic peculiarities that may not be evident from the colors alone.  Finally, following the approach of \cite{burg07} and \cite{burg10}, we form combination templates from the standards to assess whether any of our objects might be best fit as unresolved binaries. For spectral comparison to standard L and T dwarfs we used $\chi^2$ minimization over the 0.95-1.35 $\mu$m wavelength range. To assess candidate binarity we compare our spectra to combinations of L and/or T dwarf doubles over the entire 0.8-2.5 $\mu$m range, as detailed in Section 4.3. Table \ref{tab:results} lists the determined spectral types, the characteristics of each object, and the peculiarities of our objects determined from both colors and a detailed analysis of their spectra.  All of our spectra are shown in Figure \ref{fig:spectra}.

We determined that our candidate list of 40 observed objects includes 13 M dwarfs, 26 L dwarfs, and 1 T dwarf.  Of these, 10 were previously known and suspected to be L dwarfs but did not have any published near-IR spectra. The remaining 30 are new, including the T dwarf.  Ten of the 27 L and T dwarfs are either peculiar (4) or possible unresolved binaries (6). 

\clearpage
\clearpage

\begin{deluxetable}{cccccccccc}
\rotate
\tabletypesize{\scriptsize}
\tablecolumns{10}
\tablewidth{0pt}
\tablecaption{Results from Spectroscopic Classification and Synthetic Photometry
\label{tab:results}}
\tablehead{
\colhead{2MASS ID} & \colhead{IR} & \colhead{Interpretation} & \colhead{$z$} & \colhead{$J$} & \colhead{$H$} & \colhead{$K_s$} & \colhead{Disc.} & \colhead{$>$1$\sigma$ Color} & \colhead{$>$2$\sigma$ Color} \\
\colhead{(J2000)} & \colhead{SpT} & \colhead{(from spectrum)} & \colhead{(mag)} & \colhead{(mag)} & \colhead{(mag)} & \colhead{(mag)} & \colhead{Pub.} & \colhead{Outlier} & \colhead{Outlier}  }
\startdata
 & & & & High Priority & & & & & \\
\tableline
07483864+1743329 & L5 & & ...\tablenotemark{a} &  $\cdots$ \tablenotemark{b}  &  $\cdots$   &  $\cdots$   & 1 & & \\
08095903+4434216 & L7 & & 19.25 $\pm$ 0.09 & 16.22 $\pm$ 0.05 & 14.94 $\pm$ 0.05 & 14.07 $\pm$ 0.05 & 2 & $+$ & \\
09572983+4624177 & L6 & & 19.78 $\pm$ 0.08 & 16.62 $\pm$ 0.06 & 15.39 $\pm$ 0.06 & 14.59 $\pm$ 0.06 & 7 & $+$ & \\
10020752+1358556 & L7 pec & L7+T8? & 20.49 $\pm$ 0.10 & 17.72 $\pm$ 0.08 & 16.56 $\pm$ 0.06 & 15.78 $\pm$ 0.07 & & & \\
11193254-1137466 & L7 red & young & 20.69 $\pm$ 0.30 & 17.23 $\pm$ 0.12 & 15.70 $\pm$ 0.08 & 14.60 $\pm$ 0.10 & &  & $+$ \\
11260310+4819256 & L5 & & 20.07 $\pm$ 0.11 & 17.20 $\pm$ 0.07 & 16.14 $\pm$ 0.06 & 15.44 $\pm$ 0.06 & & & \\
13043568+1542521 & T0 pec & L6+T6? & 20.50 $\pm$ 0.09 & 17.24 $\pm$ 0.08 & 16.37 $\pm$ 0.06 & 15.76 $\pm$ 0.08 & &  & \\
13431670+3945087 & L5 & & 19.02 $\pm$ 0.10 & 16.02 $\pm$ 0.08 & 14.84 $\pm$ 0.07 & 14.08 $\pm$ 0.07 & 3 & & \\
14025564+0800553 & T2 pec & L8+T5? & 20.30 $\pm$ 0.09 & 17.31 $\pm$ 0.07 & 16.51 $\pm$ 0.04 & 16.01 $\pm$ 0.06 & 5 & $+$ &  \\
16005759+3021571 & L5 & & 20.25 $\pm$ 0.10 & 17.33 $\pm$ 0.07 & 16.30 $\pm$ 0.06 & 15.66 $\pm$ 0.07 & & & \\
16094569+1426422 & L4 & & 19.70 $\pm$ 0.08 & 16.92 $\pm$ 0.06 & 15.93 $\pm$ 0.06 & 15.28 $\pm$ 0.06 & & & \\
16091143+2116584 & L2 & & 20.02 $\pm$ 0.10 & 17.19 $\pm$ 0.09 & 16.25 $\pm$ 0.08 & 15.62 $\pm$ 0.09 & 1 & & \\
16135698+4019158 & L5 red & old/dusty & 20.29 $\pm$ 0.13 & 17.46 $\pm$ 0.10 & 16.30 $\pm$ 0.06 & 15.54 $\pm$ 0.10 & & $+$ & \\
16470847+5120088 & M9 & & 19.58 $\pm$ 0.10 & 17.10 $\pm$ 0.05 & 16.36 $\pm$ 0.06 & 15.91 $\pm$ 0.06 & & & \\
16592987+2055298 & M9 & & 20.14 $\pm$ 0.13 & 17.75 $\pm$ 0.08 & 17.06 $\pm$ 0.06 & 16.58 $\pm$ 0.07 & & & \\
17081563+2557474 & L5 red & young & 20.00 $\pm$ 0.08 & 16.92 $\pm$ 0.04 & 15.68 $\pm$ 0.05 & 14.84 $\pm$ 0.04 & & $+$ &  \\
17161258+4125143 & L4 & & 19.77 $\pm$ 0.16 & 16.82 $\pm$ 0.09 & 15.83 $\pm$ 0.09 & 15.16 $\pm$ 0.08 & & & \\
17373467+5953434 & L5 pec & L4+T5? & 20.80 $\pm$ 0.14 & 17.75 $\pm$ 0.08 & 16.97 $\pm$ 0.07 & 16.52 $\pm$ 0.07 & 6 & & $-$ \\
21203483-0747378 & L2 & & 19.97 $\pm$ 0.08 & 17.17 $\pm$ 0.06 & 16.24 $\pm$ 0.05 & 15.68 $\pm$ 0.05 & 6 & & \\
21243864+1849263 & L9 & & 20.15 $\pm$ 0.09 & 17.27 $\pm$ 0.05 & 16.18 $\pm$ 0.05 & 15.62 $\pm$ 0.06 & & & \\
22153705+2110554 & T1 pec & T0+T2? & 19.48 $\pm$ 0.07 & 16.22 $\pm$ 0.07 & 15.45 $\pm$ 0.06 & 15.09 $\pm$ 0.05 & &  & \\
23322678+1234530 & T0 pec & L5+T5? & 19.66 $\pm$ 0.10 & 16.92 $\pm$ 0.08 & 16.20 $\pm$ 0.07 & 15.76 $\pm$ 0.06 & & & $-$ \\
\tableline
 & & & & Lower Priority & & & & & \\
\tableline
16110632+0025469 & M9 & & ...\tablenotemark{a} &  $\cdots$ \tablenotemark{b}  & $\cdots$   &  $\cdots$   & 4 & & \\
16231308+3950419 & L3 & & 20.16 $\pm$ 0.08 & 17.35 $\pm$ 0.06 & 16.42 $\pm$ 0.06 & 15.81 $\pm$ 0.05 & & & \\
16242936+1251451 & M9 & & 19.20 $\pm$ 0.06 & 16.68 $\pm$ 0.06 & 15.88 $\pm$ 0.04 & 15.39 $\pm$ 0.06 & 5 & & \\
16304999+0051010 & L2 & & 18.40 $\pm$ 0.11 & 15.64 $\pm$ 0.07 & 14.74 $\pm$ 0.06 & 14.11 $\pm$ 0.07 & & & \\
16322360+2839567 & L1 & & 19.90 $\pm$ 0.12 & 17.30 $\pm$ 0.12 & 16.50 $\pm$ 0.10 & 16.00 $\pm$ 0.10 & & & \\
16360752+2336011 & L1 & & 19.61 $\pm$ 0.08 & 17.00 $\pm$ 0.07 & 16.16 $\pm$ 0.06 & 15.65 $\pm$ 0.06 & 5 & & \\
16370238+2520386 & L4 & & 19.95 $\pm$ 0.10 & 17.10 $\pm$ 0.08 & 16.18 $\pm$ 0.07 & 15.61 $\pm$ 0.06 & & $-$ & \\
16403870+5215505 & M9 & & 20.37 $\pm$ 0.09 & 17.97 $\pm$ 0.11 & 17.19 $\pm$ 0.08 & 16.72 $\pm$ 0.11 & & & \\
16410015+1335591 & L2 & & 20.27 $\pm$ 0.08 & 17.48 $\pm$ 0.14 & 16.56 $\pm$ 0.10 & 15.97 $\pm$ 0.14 & & & \\
17145224+2439024 & M9 & & 19.17 $\pm$ 0.12 & 16.67 $\pm$ 0.08 & 15.86 $\pm$ 0.07 & 15.35 $\pm$ 0.07 & & & \\
17251557+6405005 & L2 pec & blue & 20.10 $\pm$ 0.13 & 17.40 $\pm$ 0.08 & 16.66 $\pm$ 0.07 & 16.20 $\pm$ 0.07 & 1 & & $-$  \\
21050130-0533505 & M7 & & 18.36 $\pm$ 0.06 & 16.32 $\pm$ 0.06 & 15.65 $\pm$ 0.06 & 15.28 $\pm$ 0.06 & & & \\
21111559-0543437 & M9 & & 18.94 $\pm$ 0.09 & 16.46 $\pm$ 0.08 & 15.67 $\pm$ 0.08 & 15.17 $\pm$ 0.07 & & & \\
21115335-0644172 & M9 & & 19.83 $\pm$ 0.08 & 17.38 $\pm$ 0.06 & 16.60 $\pm$ 0.06 & 16.06 $\pm$ 0.07 & & & \\
21392224+1124323 & M8 & & 19.74 $\pm$ 0.13 & 17.54 $\pm$ 0.10 & 16.81 $\pm$ 0.08 & 16.39 $\pm$ 0.10 & & & \\
22483513+1301453 & M9 & & 19.61 $\pm$ 0.12 & 17.14 $\pm$ 0.09 & 16.38 $\pm$ 0.07 & 15.92 $\pm$ 0.08 & & & \\
23023319-0935188 & M7 & & 19.63 $\pm$ 0.10 & 17.41 $\pm$ 0.10 & 16.75 $\pm$ 0.07 & 16.36 $\pm$ 0.10 & & & \\
23443744-0855075 & M9 & & 19.43 $\pm$ 0.10 & 16.96 $\pm$ 0.06 & 16.24 $\pm$ 0.04 & 15.79 $\pm$ 0.05 & & & \\
\enddata
\tablecomments{Determination of color outliers came from comparing synthetic $J-K_s$ colors to average $J-K_s$ colors for M8-M9 and T0-T8 spectral types from \cite{faher09} and for L0-L9 spectral types from \cite{fah13}. Positives and negatives indicated whether the object was above or below the average, respectively. $^1$\cite{zhang09}, $^2$\cite{knapp04}, $^3$\cite{kirk00}, $^4$\cite{schmidt10}, $^5$\cite{chiu06}, $^6$\cite{geissler11}, $^7$\cite{luh14}.}
\tablenotetext{a}{\footnotesize FIRE spectra do not cover the entire SDSS $z$-band.}
\tablenotetext{b}{\footnotesize Target and standard observations were taken with different gain settings so individual JHK$_s$ magnitudes are not reported.}
\end{deluxetable}

\clearpage

\begin{figure}[!h]
\centering
\includegraphics[scale=0.65]{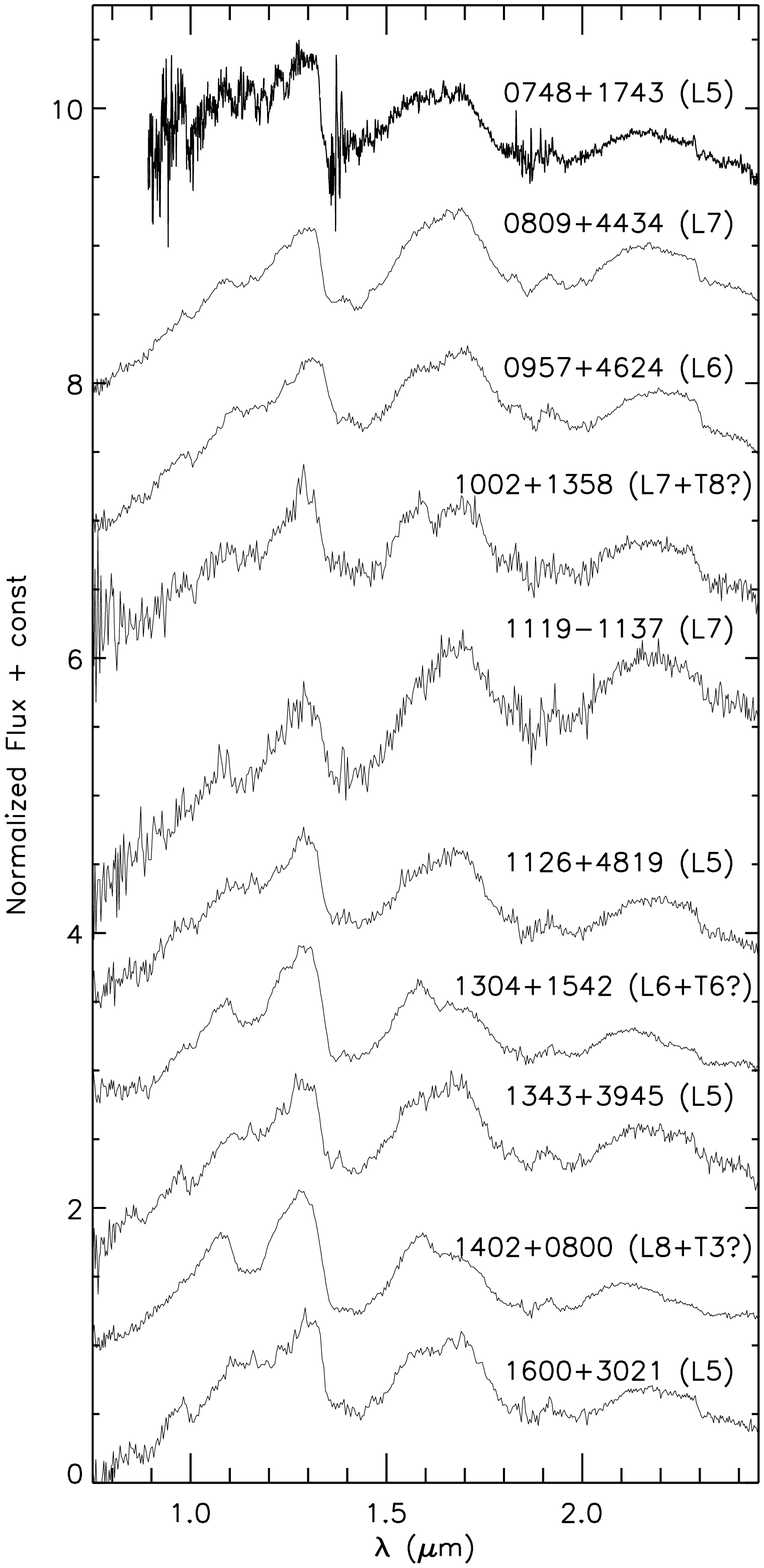}
\caption{FIRE (0748+1743, 1611+0025; $R \sim$400) and SpeX (all the rest; $R \sim$75-150) spectra of all of our reported ultra-cool dwarfs in order of right ascension. Spectral types are given in parentheses.}
\label{fig:spectra}
\end{figure}

\renewcommand{\thefigure}{\arabic{figure} (cont)}

\begin{figure}[!h]
\centering
\includegraphics[scale=0.65]{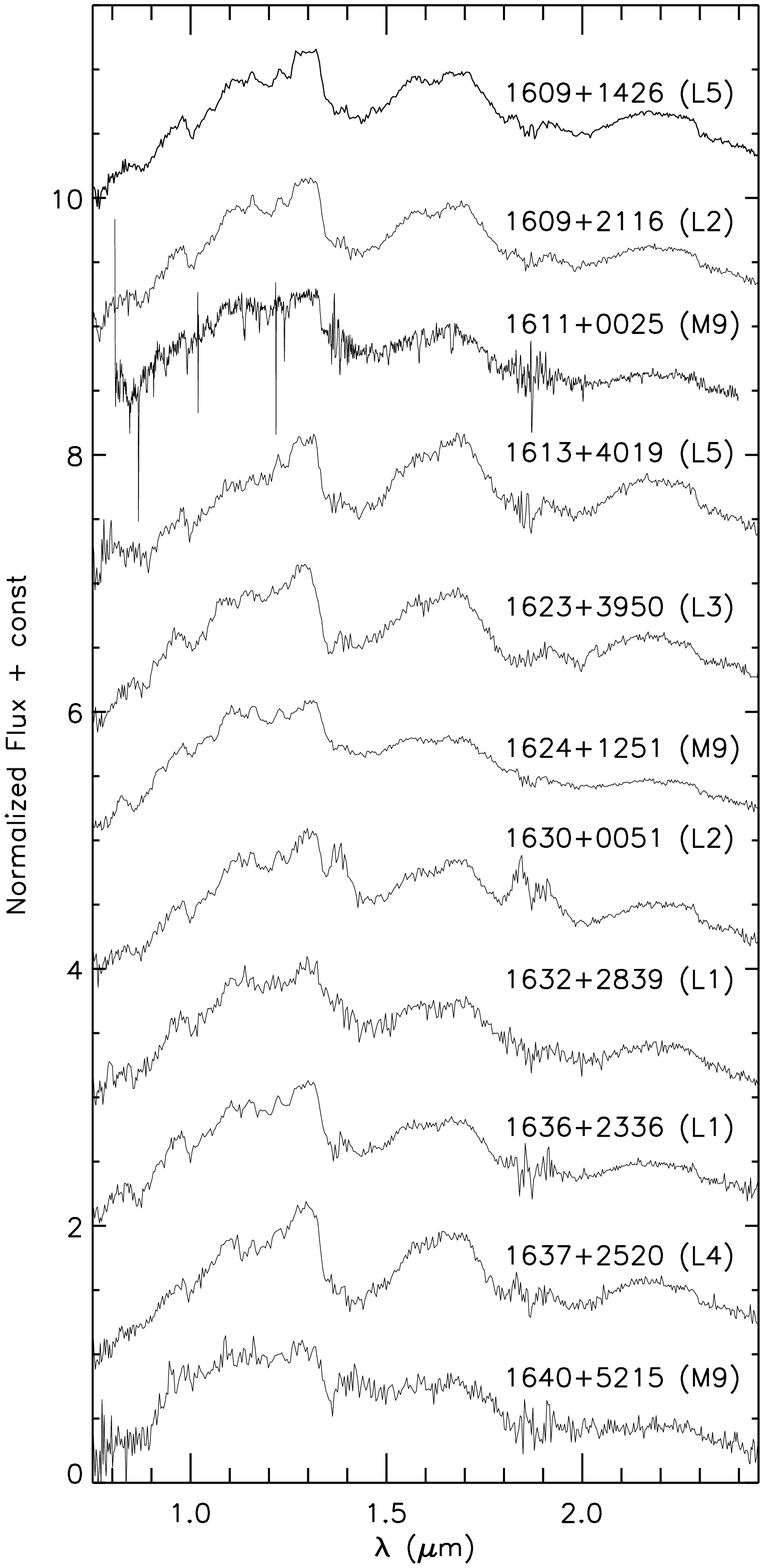}
\setcounter{figure}{2}
\caption{}
\end{figure}

\begin{figure}[!h]
\centering
\includegraphics[scale=0.65]{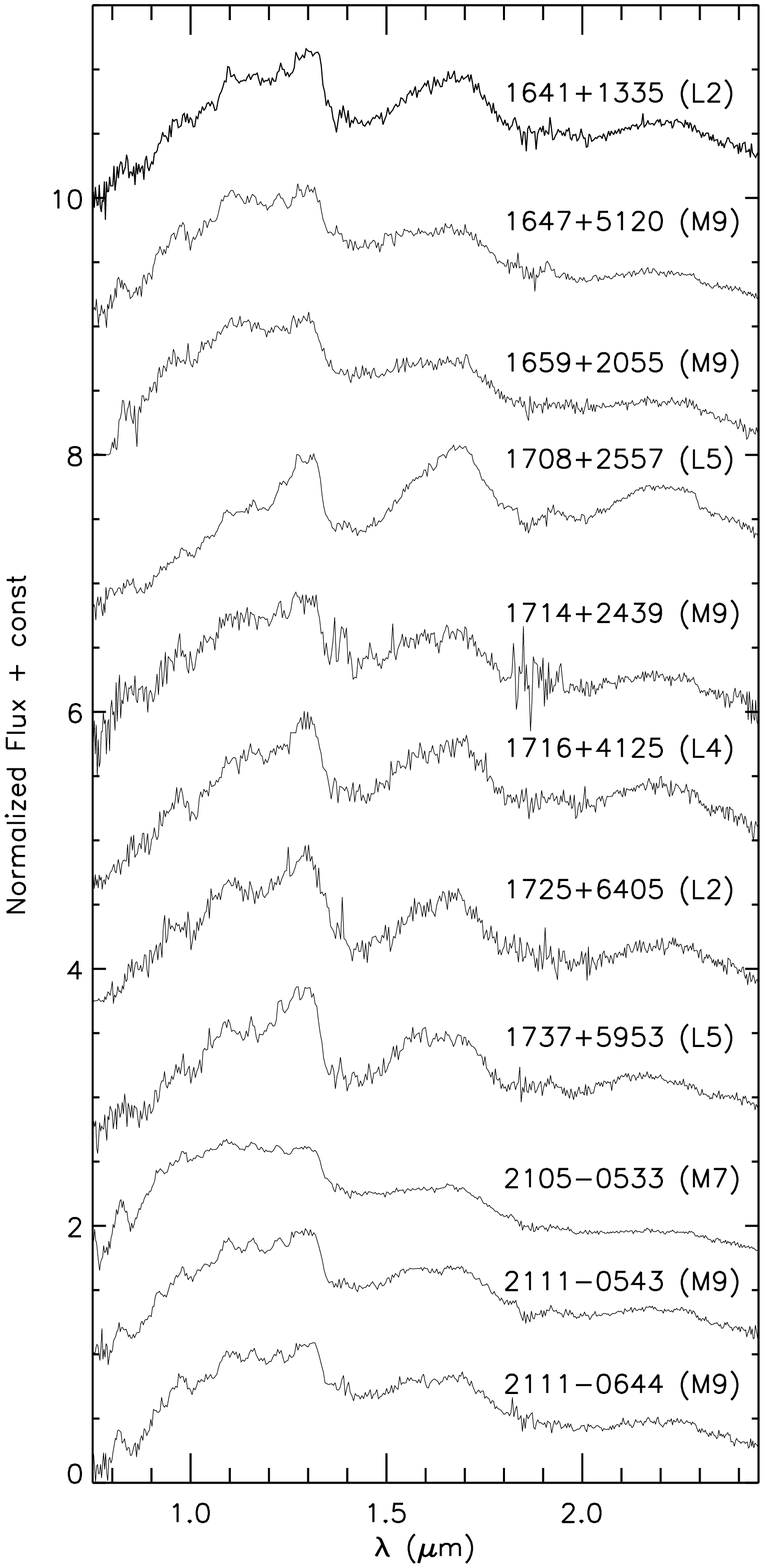}
\setcounter{figure}{2}
\caption{}
\end{figure}

\begin{figure}[!h]
\centering
\includegraphics[scale=0.65]{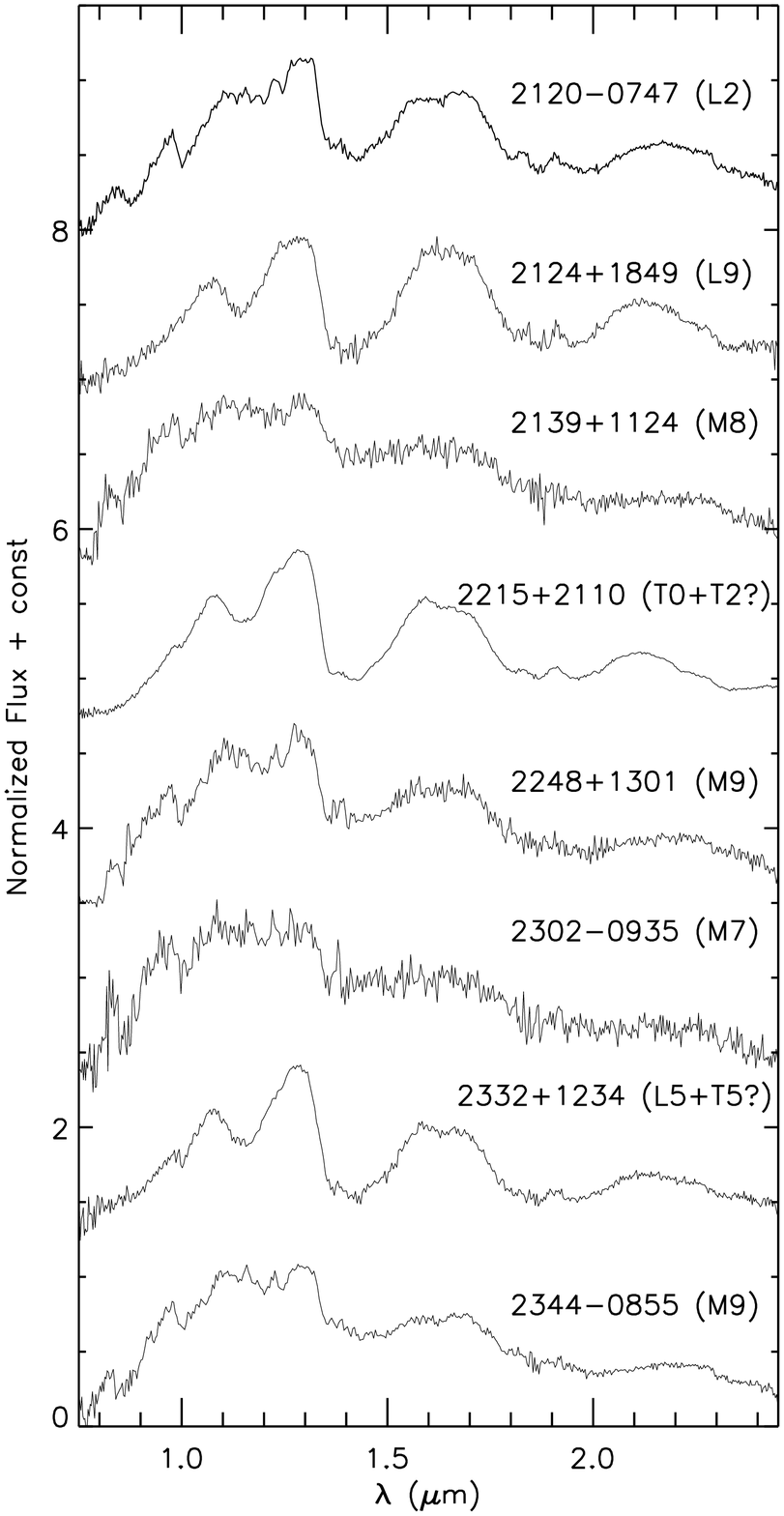}
\setcounter{figure}{2}
\caption{}
\end{figure}

\renewcommand{\thefigure}{\arabic{figure}}

\clearpage

The newly classified M, L, and T dwarfs are plotted on the $z-J$ vs. $J-K_s$ color-color diagram in Figure \ref{fig:ccdiag_paper}a, where we have used the synthetic colors integrated from the spectra.  We find that in a few cases the synthetic $z-J$ colors are bluer than 2.5 mag.  As discussed in Section 3.3, this is likely the result of flux over-estimation bias for these faint targets, mostly in the 2MASS $J$-band. We discuss the normal, peculiar, and candidate binary ultra-cool dwarfs in our sample below.

A handful of objects have synthetic colors that are bluer than the $z-J$ = 2.5 mag color selection criterion.  The SDSS and 2MASS photometry had suggested that they were redder than $z-J$ = 2.5 mag. However, their photometric SNRs from 2MASS and/or SDSS were low (see \S 3.3), and the synthetic photometry indicates that they are in fact bluer.

\subsection{Normal Ultra-cool Dwarfs}
We classify 17 of our candidates as normal L dwarfs, i.e., they do not have any readily apparent peculiarities based on their comparison to SpeX spectral standards. These objects are presented as black upward-triangles in Figure \ref{fig:ccdiag_paper}a. We find that a further 13 candidates are M7--M9 dwarfs.  These were included in our program likely because the $i-z$ and $z-J$ colors of late-M dwarfs are close to the limits of our color selection criteria (Section~\ref{sec:selection}), and because they may have been subject to flux-overestimation bias at $J$ band (Section~\ref{sec:synphot}). 

\subsection{Peculiar L Dwarfs}
Various absorption features in the near-infrared are gravity-sensitive, hence, the low-gravity of young brown dwarfs will result in line strengths that differ from those in older objects (e.g., \citealp{lucas01, gorlova03, mcgovern04, allers07, lodieu08, rice10, allers13}). Some of these features include the Na I (1.138 and 1.141 $\mu$m) and K I (1.169 and 1.178 $\mu$m, 1.244 and 1.253 $\mu$m) doublets, FeH (bandheads at 0.990 $\mu$m and 1.194 $\mu$m), and VO (1.05--1.08 $\mu$m and 1.17--1.22 $\mu$m).  Alkali lines are weaker at low gravity because of decreased pressure broadening. In low-resolution spectra these lines are often blended with other molecular features so we can not obtain accurate measurements of their strengths. Metal hydride molecular features are also weaker at low gravity because of decreased opacity from these refractory species, while VO bands are stronger (see, e.g., \citealp{kirk06}).  The 1.17um VO band is not used as a gravity indicator at low resolution because if is blended with K I, FeH and H$_2$O \citep{allers13}. Collision-induced absorption from molecular hydrogen (H$_{2}$ CIA) also changes as a function of gravity, with lower collision rates in low-gravity objects imparting a triangular shape to the $H$ band. 

Several prior analyses have introduced broad-band measures to discern low-gravity from field-gravity objects.  \cite{allers13} design several near-IR indices to measure the changing strengths of FeH, VO, and K I absorption and the slope of the $H$-band continuum as a function of gravity by comparing $\sim$1-100 Myr M5-L7 members of young moving associations with field dwarfs.  \cite{cant13} analyzed M9-L0 dwarfs to design an $H_2$($K$) index that measures the contribution of $H_2$ CIA on the slope of the $K$-band continuum; \cite{schn14} expand this index to the L dwarfs.  Indices have the potential to offer a quantitative gravity classification, analogous to spectral classification.  However, index measures depend on the spectral resolution of the data used to calibrate them, and our spectra are sufficiently distinct from those used in prior studies.  In addition, most of the indices do not extend into the late-L dwarfs, and so are inadequate to classify some of our most interesting objects.  Therefore, we do not adopt spectral indices as a default gravity classification scheme.  However, we do check for consistency with applicable spectral indices whenever we note peculiarities in the spectra of our L and T candidates.

We note that some of the spectral features, in particular the strength of the FeH bands, the peakiness of the $H$-band continuum, and the redness of the near-IR SED, may also be attributable to high atmospheric dust content or thicker clouds, as discussed in \citet{loop08b} and \citet{allers13}.  High dust content itself may be linked to low gravity, so a clear distinction may not always be possible, especially at low spectral resolution. 

Our assessment of peculiarity is based on two factors: (1) the deviation from the median J-Ks colors for objects of the same spectral type, with $>$2$\sigma$ outliers considered peculiar, or (2) high spectral similarity to objects that have previously been classified as peculiar.  In two cases below (Sections 4.2.2--4.2.3), we find similarities to the spectra of objects previously classified as peculiar because of being young and/or dusty. In the remaining two cases (Sections 4.2.1 and 4.2.4) the assessment of peculiarity is based on the comparison to spectra of previously classified peculiar objects as well as the $J-K_s$ colors. 

\subsubsection{2MASS J11193254$-$1137466 (L7)}
The most interesting object uncovered by our cross-correlation is 2M 1119$-$1137. This object is one of the reddest objects published to date with a synthetic $J-K_{s} =$ 2.62 $\pm$ 0.15 mag. Only the L7 dwarfs PSO J318.5338$-$22.8603 \citep{liu13} and ULAS J222711$-$004547 \citep{maroc14} among free-floating brown dwarfs are known to be redder. From its low-resolution spectrum (Figure \ref{fig:pec}), we classify this object as an L7. The low signal-to-noise prevents us from unambiguously determining if this object has low gravity. The peak of the $H$-band continuum --- thought to be sharpened at low surface gravity (e.g., \citealp{lucas01, allers13}) --- is not very sharp.  We measured the $H$-cont index of \cite{allers13} and found a value of 0.907, which is 1.5$\sigma$ above the medan for L7 dwarfs, and similar to the $H$-cont indices of low gravity objects.  The authors note that very red L dwarfs with no youth signatures can still exhibit a triangular $H$-band shapes and similarly high $H$-cont indices.  In summary, the $H$-cont index of 2M 1119$-$1137 is consistent with it being a low-gravity object, but we can not conclude from the index alone that it is definitely young.

\begin{figure}[h]
\centering
\includegraphics[scale=0.7]{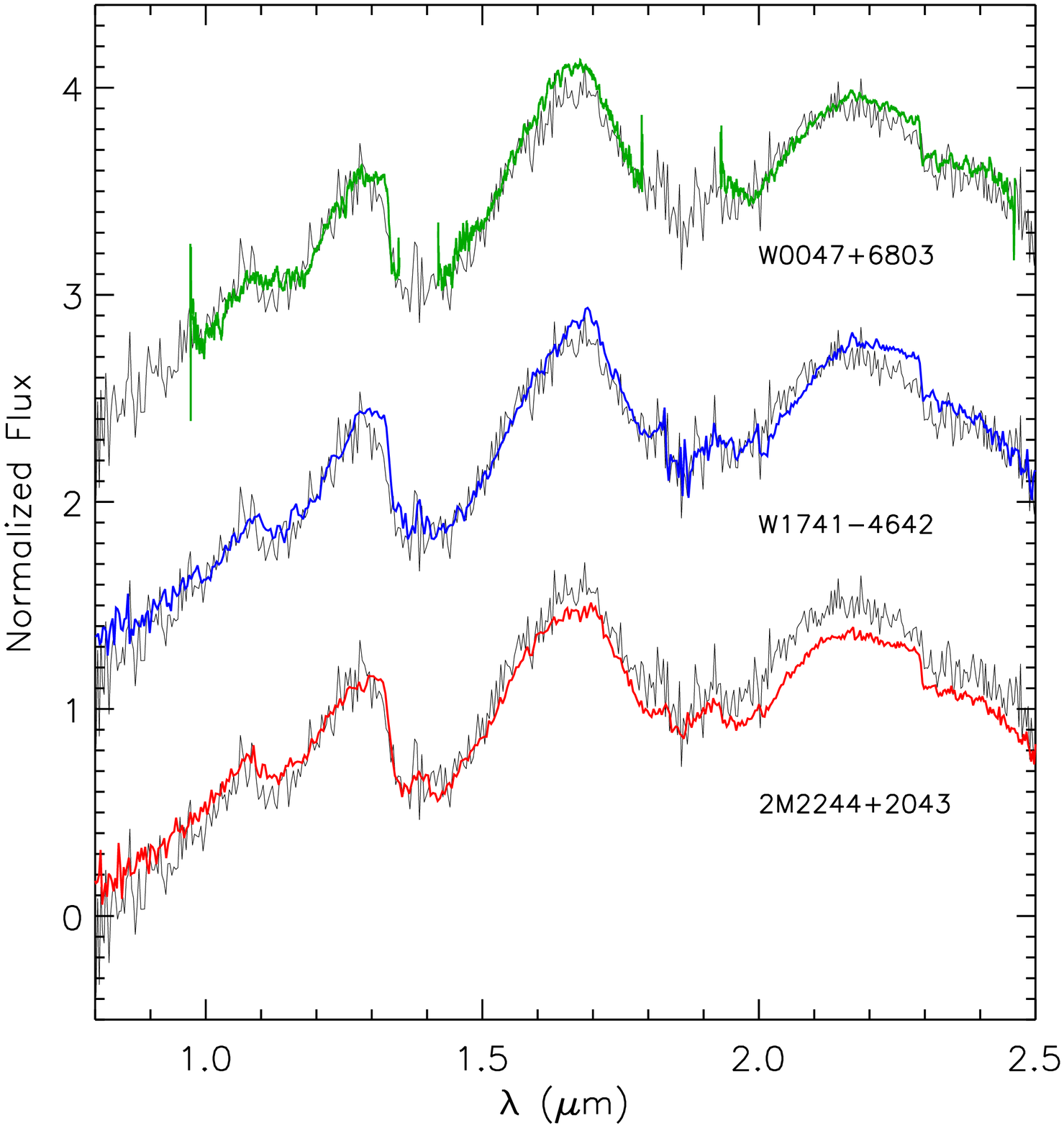}
\caption{A comparison of the SpeX prism spectrum of 2M 1119$-$1137 (black) with low-resolution spectra of other young L7--L7.5 dwarfs: WISEP J004701.06+680352.1 (L7.5 (pec), \citealp{gizis12}), WISE J174102.78$-$464225.5 (L7 (pec), \citealp{schn14}) and 2MASSW J224431.67+204343.3 (L7.5 (pec), \citealp{kirk10}).}
\label{fig:1119}
\end{figure}

In Figure \ref{fig:1119} we compare 2M 1119$-$1137 to the known very low-gravity dwarfs 2MASSW J224431.67+204343.3 (L7.5; \citealp{loop08}), WISE J174102.78$-$464225.5 (L7; \citealp{schn14}), and WISEP J004701.06+680352.1 (L7.5; \citealp{gizis12}). We see that 2M1119$-$1137 most closely matches W0047+6803 and also matches the redness of W1741$-$4642 but has a less peaked $H$-band and a shallower slope in the $K$-band. Although it is slightly redder, the shape of the $H$- and $K$-band of 2M 1119$-$1137 also matches that of 2M 2244+2043. The agreement with the spectra of other young L7--L7.5 dwarfs also indicate that 2M 1119$-$1137 may be young.  A decisive classification will require higher-SNR and/or higher-resolution spectra than we presently have.

Further evidence that 2M 1119$-$1137 may be young comes from its proper motion and photometric distance.  By comparing the 2MASS and AllWISE positions, we estimate an annual proper motion of $-$155 $\pm$ 20 mas in right ascension and $-$101 $\pm$ 17 mas in declination.  Given a $K_s$ absolute magnitude of 12.6 $\pm$ 0.4 mag for young L7 dwarfs or 12.5 $\pm$ 0.4 mag for field-age L7 dwarfs (calculated from the empirically determined $L_{\rm bol}$-SpT relationship and $K_s$ bolometric corrections from \citealp{filip15}), the photometric parallax of 2M 1119$-$1137 is 40 $\pm$ 12 mas or 38 $\pm$ 12 mas.  The BANYAN II space motion estimation algorithm \citep{malo13,gagne14} gives 2M 1119$-$1137 between 39\% and 69\% probability of being a TW Hydrae moving group member, depending on whether an arbitrary age or a $<$1 Gyr age is chosen as an input prior with the respective photometric parallax estimates.  Confirmation of the association with the TW Hydrae group will require radial velocity and trigonometric parallax measurements.

Should 2M 1119$-$1137 be confirmed as a member of the 7--13 Myr \citep{bell15} TW Hydrae association \citep{webb99}, it will be its coolest and lowest-mass (5--6 $M_{\rm Jup}$, based on evolutionary models by \citealp{allard12}) free-floating member.  Only the planetary-mass companion 2M 1207b \citep{chauv04,chauv05} is likely cooler.

\subsubsection{2MASS J17081563+2557474 (L5)}
This object is determined to be a young L5 brown dwarf based on the decreased absorption of K I and FeH and the increased absorption of H$_{2}$O in the $J$-band. Calculations of the spectral indices from \cite{allers13} and \cite{schn14} also suggest that this object is a low gravity brown dwarf. As seen in Figure \ref{fig:pec}, the strengths of the gravity sensitive features in the $J$-band and the shape of the $H$-band are more similar to the young L5 2MASS J23174712-4838501 \citep{kirk10}, although the observed spectrum is still slightly redder than the comparison spectrum.

\subsubsection{2MASS J16135698+4019158 (L5)}
\begin{figure}[h]
\centering
\includegraphics[scale=0.7]{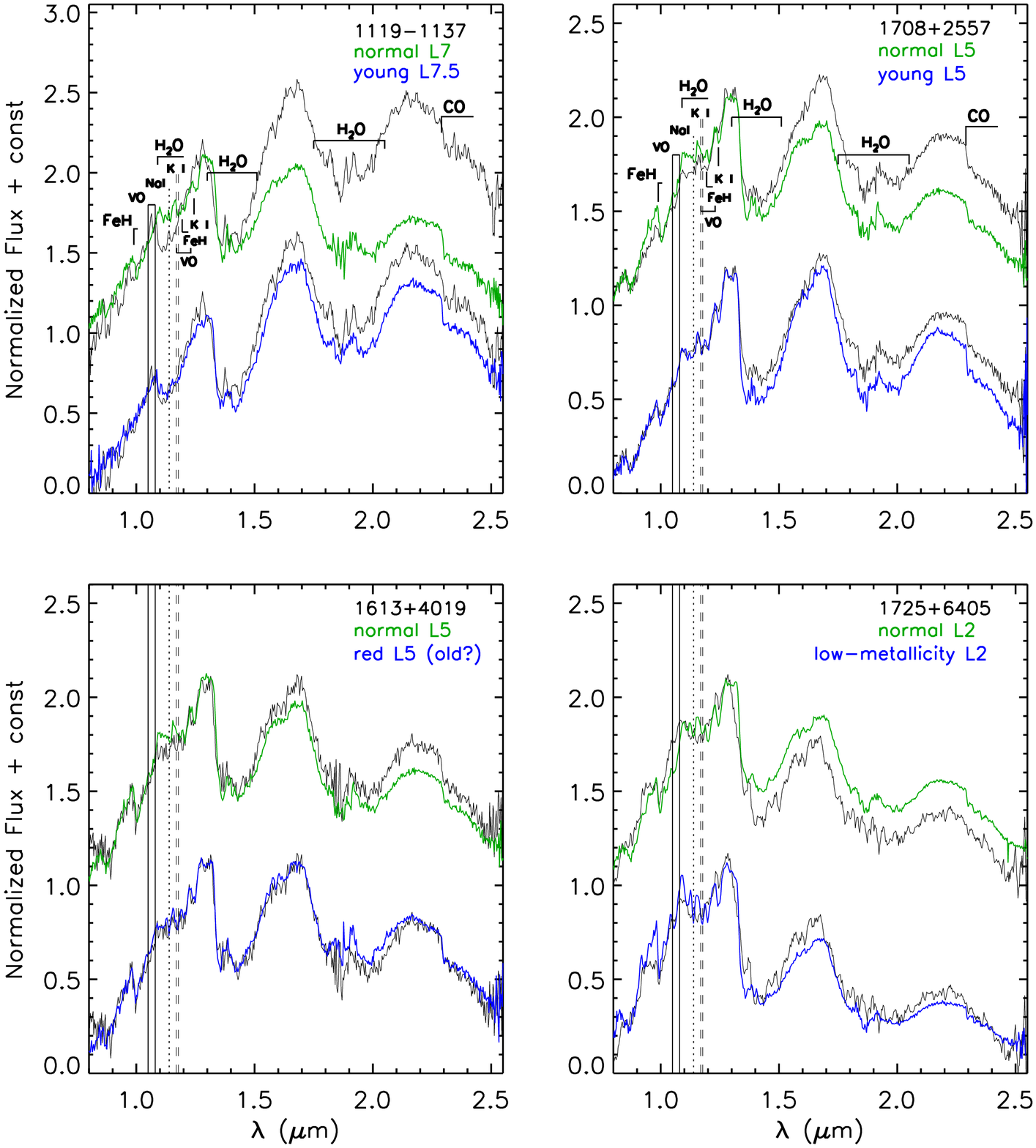}
\caption{\small Spectra of the four peculiar objects identified in this work.  In each case, the spectrum of the candidate is compared to the spectrum of a normal object of the same spectral type, and to the spectrum of a peculiar object of the nearest spectral type. The comparison spectra from left to right and top to bottom are  L7 (2MASS J0028208+224905; \citealp{burg10}) and L7.5 young (2MASS J22443167+2043433; \citealp{loop08}); L5 (2MASS J01550354+0950003; \citealp{burg10}) and L5 pec (2MASS J23174712$-$4838501; \citealp{kirk10}); L5 (2MASS J01550354+0950003; \citealp{burg10}) and L5 pec (2MASS J23512200+3010540; \citealp{kirk10}); L2 (2MASS J13054019$-$2541059; \citealp{burg07b}) and L2 pec (2MASS J14313097+1436539; \citealp{shepp09}).}
\label{fig:pec}
\end{figure}

While this object is peculiarly red, it does not exhibit the features of a low-gravity object. As seen in Figure \ref{fig:pec}, the object has normal absorption strengths, aside from H$_{2}$O, and is more similar to the red L5 dwarf 2M 2351+3010 published in \cite{kirk10}. There is also strong FeH 

\clearpage

\clearpage

\noindent absorption. The authors speculate that 2M 2351+3010 is actually an older object that simply has a higher dust content. Since our object seems very similar in nature, we adopt this explanation as well.

\subsubsection{2MASS J17251557+6405005 (L2)}
2M 1725+6405 is a peculiarly blue L2 dwarf (Fig. \ref{fig:pec}). This object was found in our cross-correlation but it was not part of our high-priority sample. Peculiarly blue L dwarfs have often been classified as metal-poor (e.g., \citealp{burg03,burg04b}), with their blue near-IR colors dictated by increasingly strong collision-induced hydrogen absorption over 1.5--2.5~$\micron$.  Metal-poor L dwarfs, or L subdwarfs, also show strong metal-hydride absorption.  However, the FeH Wing-Ford band at 0.99 $\mu$m in 2M 1725+6405 is weak compared to the standard, which suggests that the 2M 1725+6405 is blue likely because it is unusually dust-poor.

It is also possible that 2M 1725+6405 may be an unresolved L + T dwarf binary, with the $J$ band flux enhanced by the T dwarf component.  We consider unresolved binarity in the next Section (\ref{sec:binaries}).  Unlike all of the candidate binaries discussed in Section~\ref{sec:binaries}, we actually do not find a better binary template fit for 2M 1725+6405.  We therefore conclude that this L2 dwarf is intrinsically blue.

\subsection{Brown Dwarfs with Composite Spectral Types
\label{sec:binaries}}
Several of the objects show peculiarities that do not readily match those found in other individual objects.  Instead, they more closely resemble combination spectra of L and T dwarfs. \cite{burg07} and \cite{burg10} developed a technique that enables one to infer the spectral types of the individual components of a candidate unresolved binary by a goodness-of-fit comparison to a library of spectral template combinations.  We adopt this technique in a simple form, by creating combination templates from the set of single L and T dwarf standards from the SpeX Prism Library.  Unlike \citet{burg10}, we do not create a large list of templates built on the entire population of L and T dwarfs with available SpeX spectra.  Nonetheless, we find that our simple approach gives sufficient indication whether a brown dwarf displays a composite spectral signature, and produces approximate spectral types for the components.

Our composite template spectra are constructed by normalizing all of the standard single brown dwarfs over the same wavelength range (1.2-1.3 $\mu$m; chosen because it is relatively free of absorption features), scaling them to their absolute spectral-type dependent magnitudes given by the polynomials in Table 14 of \cite{dup12}, and summing the pairs of resulting spectra. We compute the $\chi^2$ over most of the 0.8-2.5 $\mu$m region, excluding ranges of strong water absorption (1.35-1.45 and 1.8-2.0 $\mu$m). In all cases, the $\chi^{2}$ is greater than one but this is to be expected as we are only testing the fit to templates created from one object of each spectral type. We have classified an object as a likely spectral type composite -- a potential binary -- if the $\chi^2$ of the dual-template spectral fit is significantly lower than the $\chi^2$ of the single-template fit. Each of the $\chi^2$ values have been calculated over the entire 0.8-2.5 $\mu$m region, minus the water absorption bands.

In addition to template fitting, we have analyzed the spectral indices defined specifically for SpeX prism spectra in \cite{burg10} for all our binary candidates and we report the strength of their candidate binarity. We have also analyzed the SpeX prism spectral indices from \cite{bard14} but because the binary index selection criteria in that work were not designed for late-L to early-T dwarfs, we only report the results where applicable.

We note that while brown dwarfs displaying combination spectral signatures have until recently been considered to all be unresolved binaries, they can also be highly variable brown dwarfs with photospheres that display two distinct temperature components.  Recent examples include the  T1.5 dwarf 2MASS J21392676+0220226, suggested as a strong L8.5 + T3.5 spectral binary candidate by \citet{burg10}, but identified as a $J$-band variable that is unresolved in {\it HST} images \citep{rad12}, or the T dwarfs 2MASS J13243559+6358284 (T2.5) and SDSS J151114.66+060742.9 (T2), identified as binary candidates \citep{burg10, geissler11}, but that are also unresolved in {\it HST} and are variable \citep{metchev15}.  Therefore, while the objects discussed in this section are considered candidate unresolved binaries, they are also strong candidates for photometric variables.

\subsubsection{2MASS J13043568+1542521 (L6+T6?)}
This object is one of several that is best fit by a binary combination template. As seen in Figure \ref{fig:bin}, the best fit single brown dwarf (T0) does not match the features of 2M 1304+1542. The $Y$-/$J$-band ratio is lower than any of the closest standard objects and the $H$-band has a dip at $\sim$1.65 $\mu$m. The $K$-band does not have differences that are as pronounced as in the other bands though it is slightly redder than the standard object. In fitting this object with a binary template, we find that the best fit is a combination of an L6 and a T6 brown dwarf. The $Y$-/$J$-band ratio and the $K$-band flux more closely resemble the object spectrum. The contribution of the methane break in the cooler brown dwarf at 1.6 $\mu$m also reproduces the dip in the $H$-band well. Further evidence that this object is a binary comes from the analysis of spectral indices identified in \cite{burg10} and \cite{bard14}. 2M 1304+1542 satisfies four of the six binary index selection criteria given in Table 5 of \cite{burg10} and ten of the twelve selection criteria in Table 4 of \cite{bard14}, making this a strong binary candidate. 

\subsubsection{2MASS J14025564+0800553 (L8+T5?)}
The spectrum of 2M 1402+0800 also shows distinctive composite characteristics. While the $Y$-/$J$-band ratio is not significantly dissimilar from the closest single brown dwarf spectrum, the 

\clearpage

\begin{figure}[h]
\centering
\includegraphics[scale=0.8]{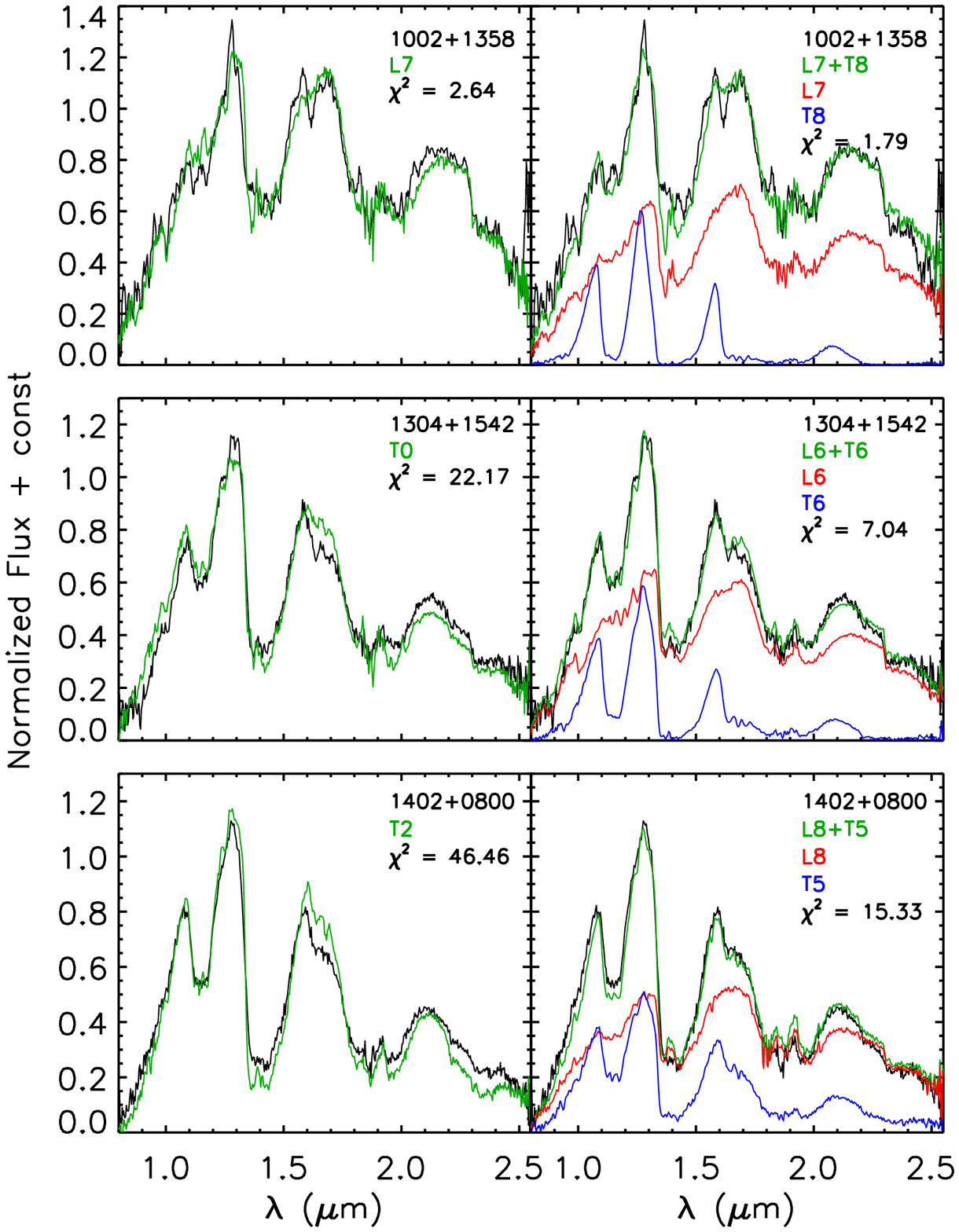}
\caption{\small Spectra of all objects identified as candidate unresolved binaries (or photometric variables). The left panels show comparisons to the spectra (in green) that fit the 0.95-1.35 $\mu$m continuum best: i.e., as done for spectral typing of the individual objects in Sections 4.1--4.2. The right panels show the two-component templates (also in green) that fit best over 0.8-2.5 $\mu$m; the individual component contributions are shown in red and blue. The quoted $\chi^2$ values are the smallest ones for, respectively, single- and binary-template fits over the entire 0.8-2.5 $\mu$m range, as done in Section 4.3.  The comparison spectra from left to right and top to bottom are:  L7 (2MASS J0028208+224905; \citealp{burg10}) and T8 (2MASS J04151954-0935066; \citealp{burg04}); T0 (2MASS J12074717+0244249; \citealp{loop07}), L6 (2MASS J10101480-0406499; \citealp{reid06}) and T6 (2MASS J16241436+0029158; \citealp{burg06b}); T2 (2MASS J12545393-0122474; \citealp{burg04}), L8 (2MASS J16322911+1904407; \citealp{burg07}) and T5 (2MASS J15031961+2525196; \citealp{burg04}).}
\label{fig:bin}
\end{figure}

\renewcommand{\thefigure}{\arabic{figure} (cont)}

\begin{figure}[h]
\centering
\includegraphics[scale=0.8]{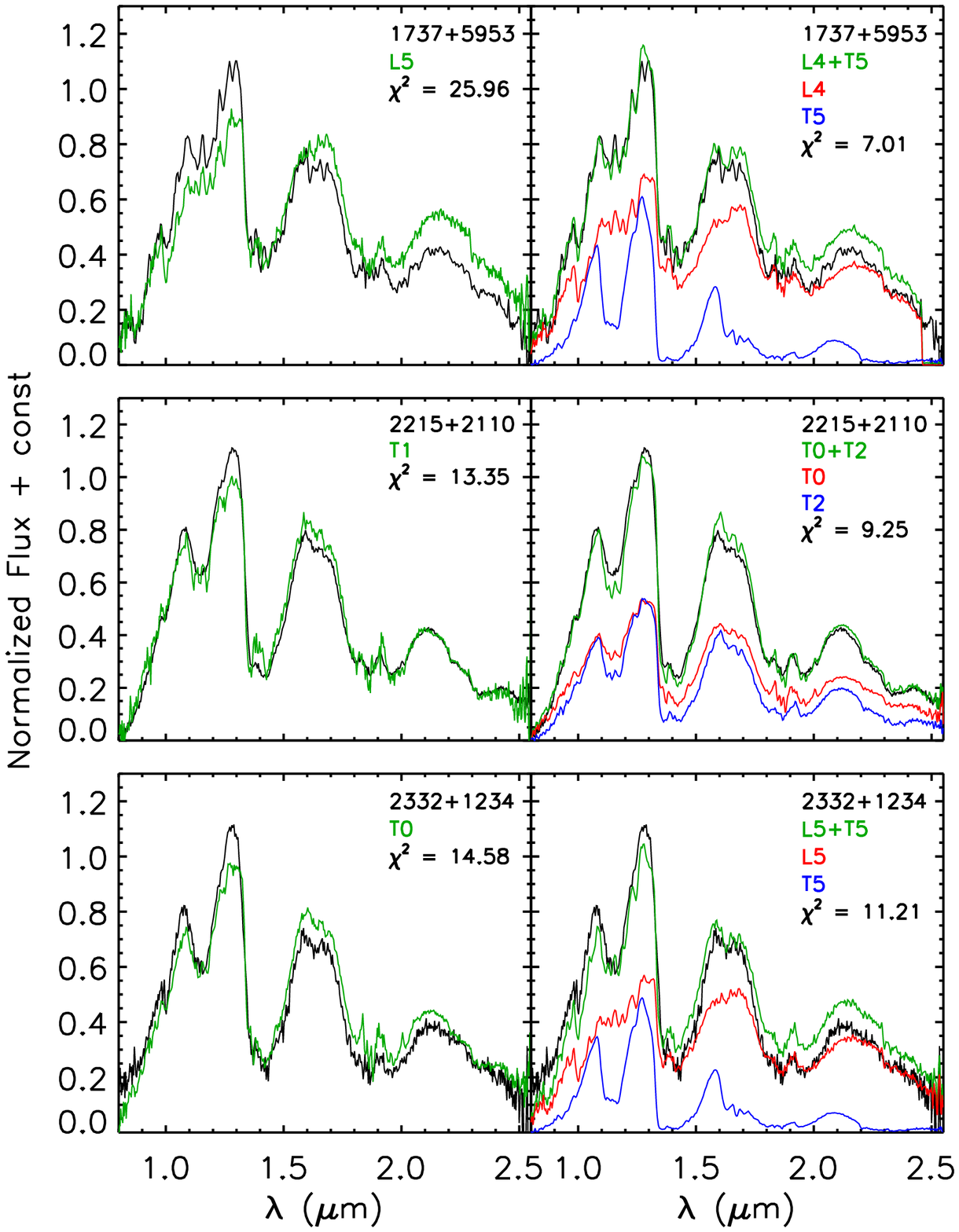}
\setcounter{figure}{5}
\caption{The comparison spectra from left to right and top to bottom are: L5 (2MASS J08350622+1953050; \citealp{chiu06}), L4 (2MASS J21580457-1550098; \citealp{kirk10}) and T5 (2MASS J15031961+2525196; \citealp{burg04}); T1 (2MASS J01514155+1244300; \citealp{burg04}), T0 (2MASS J12074717+0244249; \citealp{loop07}) and T2 (2MASS J12545393-0122474; \citealp{burg04}). L5 (2MASS J08350622+1953050; \citealp{chiu06}) and T5 (2MASS J15031961+2525196; \citealp{burg04}).}
\end{figure}

\renewcommand{\thefigure}{\arabic{figure}}

\clearpage

\noindent $H$- and $K$-bands are more similar to an L8+T5 binary. Figure \ref{fig:bin} shows that the shape and relative flux of all three 2MASS bandpasses are very well reproduced by the L/T binary template. Most importantly, the dip in the $H$-band is well reproduced by the contribution of the methane in the T dwarf. This object passes all six of the binary index selection criteria of \cite{burg10} which makes it a strong binary candidate.

\subsubsection{2MASS J17373467+5953434 (L4+T5?)}
This object is classified as having an L4+T5 composite spectrum. As seen in Figure \ref{fig:bin}, an L5 spectrum matches 2M 1737+5953 well in the $Y$- and $J$-bands but is a very poor match to the $H$- and $K$-bands. The observed spectrum shows signs of methane absorption at 1.6 and 2.2 $\mu$m which is indicative of having a T dwarf secondary component. The binary index selection criteria from \cite{burg10} were not designed for mid-L dwarfs so we analyzed the spectral indices from \cite{bard14} instead. Because this object only passes four of the twelve selection criteria from \cite{bard14}, it is only a weak binary candidate.

\subsubsection{2MASS J23322678+1234530 (L5+T5?)}
While 2M 2332+1234 is best fit in the $J$-band by a scaled T0 spectrum, the $H$- and $K$-bands clearly do not appear to belong to a T0 dwarf. The $H$-band shows evidence of methane absorption at 1.6-1.8 $\mu$m but there is less presence of CH$_4$ in the $K$-band. This points to a composite L/T spectrum similar to SDSS J151114.66+060742.9 presented in \cite{burg10}. The methane absorption features are best fit by an L5+T5 template, however, the continuum of our observed spectrum is still slightly bluer at the longer wavelengths. This object passes four of the binary index selection criteria of \cite{burg10} which makes it a strong binary candidate.

\subsubsection{2MASS J10020752+1358556 (L7+T8?)}
This object is tentatively classified as having a composite spectrum. As seen in Figure \ref{fig:bin}, 2M 1002+1358 has a large dip in flux in the $H$ band at the location of the CH$_{4}$ absorption feature that is usually present in a T dwarf, and has much more water and methane absorption in the $J$-band than a typical L dwarf. The $K$- band, however, seems to be similar to an L4--L6 dwarf. These suggest a composite spectral type. There is a much greater difference between L and T dwarfs in the $J$- and $H$-band features than there is in the $K$-band features, therefore, the $K$-band of a combined binary spectrum can look like it belongs to an L dwarf whereas the $J$- and $H$-bands will appear to have a contribution from both binary components. The large dip in $H$-band flux may also be the result of an extraneous signal in the raw spectrum of the object as it has an atypical shape compared to that of a feature usually associated with CH$_{4}$. However, the spectrum of the telluric calibration star does not exhibit the same behavior, while the feature is apparent in most of the individual spectra of this object, even if at low SNR.  This suggests that the feature may be real, even if we can not fully exclude a random variation due to noise. Analyzing the spectral indices does not shed any light on the true nature of this object as it only passes four of the twelve binary index selection criteria from \cite{bard14}, making it a weak binary candidate.

\subsubsection{2MASS J22153705+2110554 (T0+T2?)}
The T dwarf 2M 2215+2110 is a new discovery in the SDSS footprint. Some of the features in the spectrum of 2M 2215+2110 are ambiguous as to their origin. While the $J$- and $K$- bands more closely resemble an early T dwarf, the $H$- band has a clear dearth of flux. The overall shape of this band might be explained by a presence of a slightly later-type T dwarf secondary component than the primary, but the lack of flux still persists in the binary template spectrum. Several features, such as the FeH feature at 0.99 $\mu$m, do match a T0+T2 composite spectrum.  However, the H$_{2}$O + CH$_4$ absorption between 1.1--1.2 $\mu$m is much stronger in the binary composite template than in the observed spectrum. The spectral indices also do not help us with this object -- only two of the index selection criteria from \cite{burg10} are passed which makes this object a weak binary candidate. 

\section{Discussion}
Our search was aimed at discovering peculiar L or T dwarfs, with priority in this first iteration placed on unusually red objects.  Overall, we have observed and identified 10 peculiar or binary L dwarfs, 16 normal L dwarfs, one T dwarf, and 13 M dwarfs.  The latter had been mis-identified as candidate L or T dwarfs because of low-SNR photometry.

The total fraction of objects in an unbiased sample of brown dwarfs with $J-K_s$ colors $>$2$\sigma$ from the mean color at a given spectral type --- the criterion used for detecting photometrically peculiar L and T dwarfs in \cite{faher09} --- is expected to be 4.6\%.  \cite{faher09} report a somewhat larger fraction, 5.8\%, of peculiar objects among the 1268 M7--T8 dwarfs in their sample.  The small discrepancy arises from an apparent non-gaussianity of the $J-K_s$ color distribution: they have nearly twice as many red outliers than blue outliers. 

Only three of our L dwarfs are peculiarly red or dusty, and an equal number of our discoveries are in fact peculiarly blue.  While at face value this does not indicate a higher success rate in finding peculiarly red objects than in a random sample of field brown dwarfs, we have at present followed up only a small number (40) of our total candidate sample (314).  The 40 objects presented here comprise roughly equal numbers of high- (22) and low-priority (18) objects: a circumstance of weather and observational constraints.  It is possible that the larger high-priority sample (178 candidates) will reveal a higher incidence rate of unusually red objects.

We do find, however, that our present prioritization strategy reveals a larger fraction of unusual objects --- including not only peculiar L dwarfs but also candidate unresolved binaries that are not color outliers in $J-K_s$ but are unusually red in $z-J$ --- among the high-priority candidates. Eight of the 22 objects in the high-priority sample are peculiar or candidate binaries vs. two of the 18 in the low-priority sample.  The difference between the two is statistically significant at the 96\% level.  It indicates that combinations of optical and infrared colors, such as employed here, can successfully discern even moderate peculiarities in ultra-cool dwarfs. Table \ref{tab:results} summarizes the peculiarities of each object --- from spectral comparison and synthetic colors. 
 
\begin{figure}[h]
\centering
\includegraphics[scale=0.6]{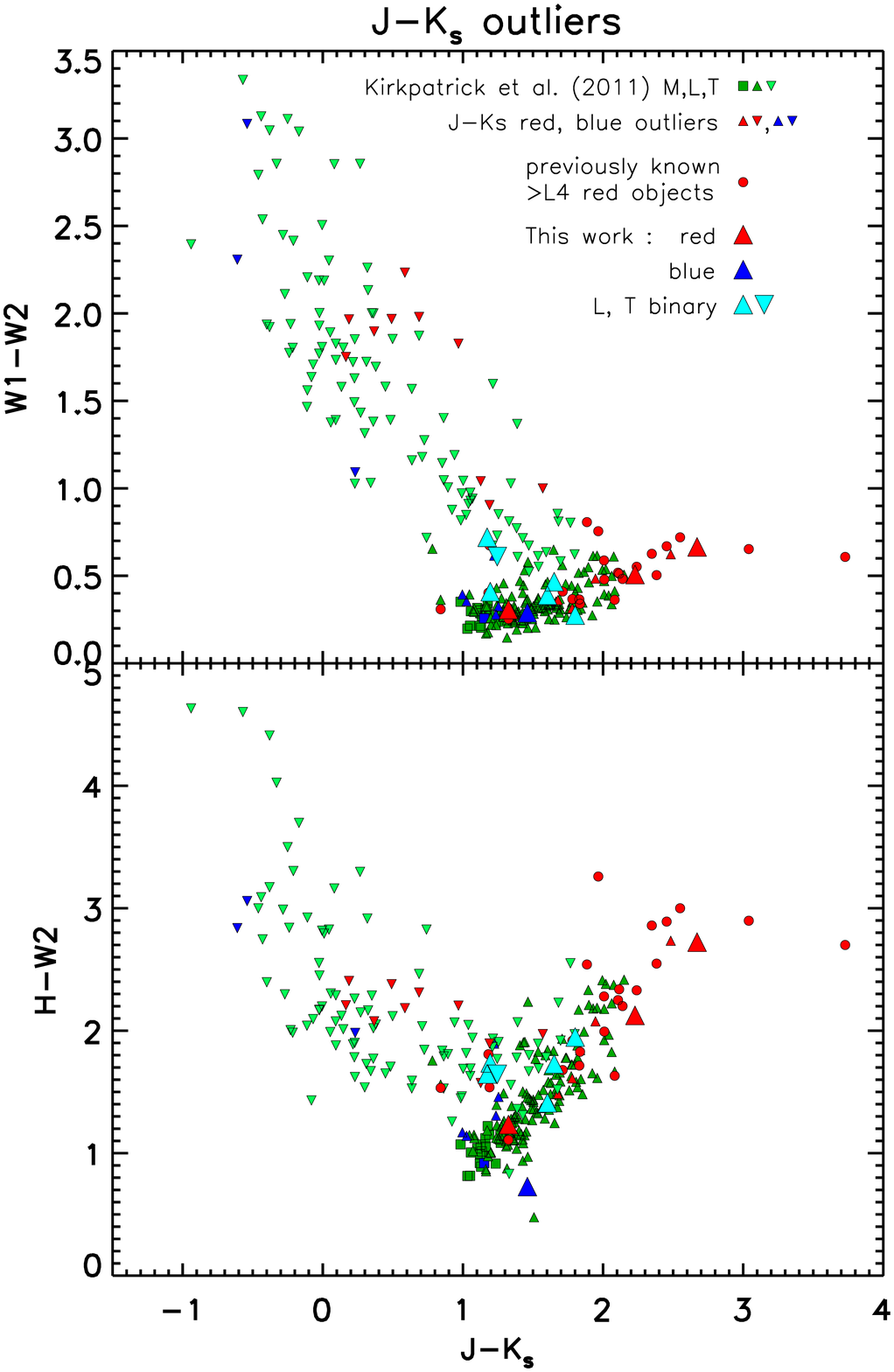}
\caption{Photometric color-color diagrams of objects from \cite{kirk11}. Upwards and downwards triangles denote L and T dwarfs, respectively.  Red symbols denote objects with $J-K_s$ colors $>$2$\sigma$ redder than the mean for their spectral type \cite{faher09,fah13}.  Blue symbols denote objects that are $>$2$\sigma$ bluer. Large symbols represent peculiar objects identified in this work. Red circles indicate the previously known red brown dwarfs with spectral types of L4 and later.}
\label{fig:jkout}
\end{figure}

Because L and T dwarfs are brighter in the 3--5 $\mu$m wavelength range, we investigated whether the $J-K_s$ color outliers also have unusual colors at these wavelengths. We find that L dwarfs with the very reddest $J-K_s$ colors are clearly distinguishable from the locus of L dwarfs on a $J-K_{s}$ vs.\ $H-W2$ and $J-K_{s}$ vs.\ $W1-W2$ diagram (Fig. \ref{fig:jkout}) mainly because of their red near-IR colors.  They stand out in their $J-K_s$ and $H-W2$ colors but not significantly in their $W1-W2$ colors. T dwarfs with peculiarly red $J-K_s$ colors are only marginally redder in $H-W2$ and $W1-W2$, and the peculiarly blue L or T dwarfs are not distinguishable from the normal population with the exception of the blue L dwarf discovered in this work (2MASS J17251557+6405005).

\section{Conclusions}
We performed a color-selected search for peculiar L and T dwarfs, focusing primarily on the peculiarly red objects, and demonstrated that with the proper selection criteria, we can identify unusual L and T dwarf candidates in large photometric surveys in the absence of spectral type information. With follow-up spectroscopy, we can verify the unusual properties and begin to discern their underlying cause. This is particularly advantageous for finding isolated objects that are analogous to the typically very red directly imaged extrasolar planets in order to study their atmospheric characteristics at higher fidelity. We had a high success rate in discovering either peculiar L dwarfs or candidate unresolved binaries in our prioritized sample, and discovered one of the reddest L dwarfs known to date. This new red L7 dwarf is a potential TW Hydrae member, and if confirmed, would make it the coolest and least massive free-floating object in the association. We note that even after many searches for T dwarfs in the SDSS and 2MASS catalogs, we still uncovered a new T dwarf among the $\sim$13\% fraction of candidates that we have spectroscopically characterized so far.  These discoveries attest to the power of simultaneous positional and color cross-correlations across photometric databases --- as performed here, in \cite{metchev08}, in \cite{geissler11}, and now enabled with the Virtual Astronomical Observatory --- over color-only searches on individual databases that are then positionally compared to other databases.   At the same time, the discovery of only a single new T dwarf in our characterized sample indicates the census of T dwarfs (132) in SDSS is nearly complete.

\clearpage

\clearpage

\bibliography{bibliography}
\bibliographystyle{apj}

{\it Facilities:} \facility{IRTF (SpeX spectrograph)}, \facility{Magellan: Baade (FIRE spectrograph)}

\acknowledgements
We would like to thank our referee, Jackie Faherty, for her insightful critique. 

This work was supported by the NASA Astrophysical Data Analysis Program through award No.\ NNX11AB18G to S.M. at Stony Brook University and by an NSERC Discovery grant to S.M. at The University of Western Ontario.  Part of the data are obtained with the Magellan-Baade 6.5m telescope under program CN2012A-54. Support for R.K. is provided from Fondecyt Reg. No. 1130140 and by the Ministry of Economy, Development, and Tourism�s Millennium Science Initiative through grant IC12009, awarded to The Millennium Institute of Astrophysics (MAS). The authors wish to recognize and acknowledge the very significant cultural role and reverence that the summit of Mauna Kea has always had within the indigenous Hawaiian community.  We are most fortunate to have the opportunity to conduct observations from this mountain. This paper includes data gathered with the 6.5 meter Magellan Telescopes located at Las Campanas Observatory, Chile. This publication makes use of data products from the Two Micron All Sky Survey, which is a joint project of the University of Massachusetts and the Infrared Processing and Analysis Center/California Institute of Technology, funded by the National Aeronautics and Space Administration and the National Science Foundation. Funding for SDSS-III has been provided by the Alfred P.\ Sloan Foundation, the Participating Institutions, the National Science Foundation, and the U.S. Department of Energy Office of Science. The SDSS-III web site is http://www.sdss3.org/.  SDSS-III is managed by the Astrophysical Research Consortium for the Participating Institutions of the SDSS-III Collaboration including the University of Arizona, the Brazilian Participation Group, Brookhaven National Laboratory, Carnegie Mellon University, University of Florida, the French Participation Group, the German Participation Group, Harvard University, the Instituto de Astrofisica de Canarias, the Michigan State/Notre Dame/JINA Participation Group, Johns Hopkins University, Lawrence Berkeley National Laboratory, Max Planck Institute for Astrophysics, Max Planck Institute for Extraterrestrial Physics, New Mexico State University, New York University, Ohio State University, Pennsylvania State University, University of Portsmouth, Princeton University, the Spanish Participation Group, University of Tokyo, University of Utah, Vanderbilt University, University of Virginia, University of Washington, and Yale University. This publication makes use of data products from the Wide-field Infrared Survey Explorer, which is a joint project of the University of California, Los Angeles, and the Jet Propulsion Laboratory/California Institute of Technology, funded by the National Aeronautics and Space Administration. This research has made use of data obtained from or software provided by the US Virtual Astronomical Observatory, which is sponsored by the National Science Foundation and the National Aeronautics and Space Administration.

\end{document}